\documentclass[aps, prb, twocolumn, longbibliography]{revtex4-2}
\usepackage{amsmath, amsfonts, amssymb}
\usepackage{graphicx, color}
\usepackage{babel}
\usepackage[dvipsnames]{xcolor}
\usepackage{xspace}
\usepackage{nicefrac, bm}
\usepackage[colorlinks, breaklinks, linkcolor=OrangeRed, citecolor=RoyalBlue, urlcolor=NavyBlue]{hyperref}
\usepackage[all]{hypcap} 

\begin{document}

\title{The covariance matrix spectrum of correlated charge insulators reveals
hidden connections to Coupled Cluster, Matrix Product, and Rokhsar-Kivelson states}

\author{Izak Snyman}
\affiliation{Mandelstam Institute for Theoretical Physics, School of Physics, University of the Witwatersrand,
Johannesburg, South Africa}
\author{Serge Florens}
\affiliation{Univ. Grenoble Alpes, CNRS, Grenoble INP, Institut Néel, 38000 Grenoble, France}

\begin{abstract}
Charge ordering induced by strong short-range repulsion in itinerant fermion systems typically follows 
a two-sites alternation pattern. However, the covariance matrix spectrum of the one-dimensional,
half-filled, spinless $t$-$V$ model reveals 
a post-Hartree-Fock picture at strong repulsion, with emergent four-site disruptions of the underlying 
staggered mean-field state.
These disruptions are captured in a thermodynamically extensive manner by a compact four-fermion
 Coupled Cluster (doubles) state (CCS). Remarkably, all properties of this state may be computed analytically 
by combinatorial means, and also derived from an exactly solvable correlated hopping Hamiltonian. 
Furthermore, this Coupled Cluster state can be re-expressed as a low-rank Matrix Product State (MPS) with 
bond dimension exactly four.
In addition, we unveil a hidden connection between this Coupled Cluster ansatz and a Rokhsar-Kivelson state (RKS), 
which is the ground state of a solvable parent quantum tetramer model. 
The broad picture that we uncover here thus provides deep connections between several core concepts of correlated fermions 
and quantum chemistry that have previously enjoyed limited synergy.
In contrast to
a recent perturbative treatment on top of Hartree-Fock 
theory,
our approach asymptotically captures the correct 
correlations in the $t$-$V$ model at small $t/V$, and remains a qualitatively accurate
approximation even outside the perturbative regime.
Our results make the case for further studies of the covariance matrix for correlated electron systems in which ground 
states have non-trivial unit-cell structure.
\end{abstract}

\maketitle
\section{Introduction}
The covariance matrix $Q_{jj'} =\big<a_j^\dagger a_{j'}\big>$ of an interacting many-body system of fermions 
$a^\dagger_j$ has a dual nature.
As it constitutes a one-body density matrix, it allows one to evaluate any {\em single-particle} 
observable~\cite{Giuliani}.
At the same time, its spectrum provides a basis-independent measure of {\em many-body} correlations~\cite{Lowdin1955}. 
Indeed, by counting the number of eigenvalues significantly different from zero or one,
one can learn how far away the state under consideration is from a non-interacting state. 
The single-particle basis that diagonalizes the $Q$-matrix is called the natural orbital basis~\cite{Davidson1972}.
Of course, when the ground state of a many-fermion system has the same translation symmetry as the
underlying lattice, the natural orbitals are momentum $k$ Bloch waves~\cite{Flensberg}, and the spectrum of $Q$ is 
nothing but the Bloch momentum occupation numbers $\left<n_k\right>$ that characterize the properties of the
emergent Fermi liquid (or Luttinger liquid in one dimension~\cite{Giamarchi2003}). However, in insulating systems like Mott
or charge insulators, the ground state has lower translational symmetry than the underlying lattice due to 
charge or spin ordering, and an interesting situation arises.
On the one hand, the symmetry-broken state is often understood in terms
of Hartree-Fock mean-field theory~\cite{Flensberg}, which yields an effective non-interacting system. 
Yet on the other hand, it is by virtue of correlation-inducing interactions that the insulating state is
achieved. What can one learn from the covariance matrix of such a system, whose natural orbitals are
now not simply Bloch waves of the underlying lattice? Existing literature does not yet provide the
answer. In this Article, we start to fill in the gap. Along the way we uncover several emergent structures
of correlated insulators that were deeply hidden from previous analytical or numerical understanding.

Quantum chemists have long understood the value of natural orbitals, when the state under
consideration is the ground state of an interacting few-electron system~\cite{Olsen2011}. More
recently, it has stimulated the interest of condensed matter physicists. Studying the spectrum of
the $Q$-matrix, it was realized that dynamical quantum impurity problems such as the ground state
\cite{He2014,Debertolis2021} and low-energy quasiparticle excitations~\cite{Snyman2023} of the Kondo model,
which are paradigmatic in quantum many-body theory, host only a handful of correlated natural
orbitals, the complement being either completely filled or empty~\cite{Bravyi2017} for all practical
purposes. When cast in the natural orbital basis, these problems thus become few-body problems
\cite{Debertolis2022}, and this structure remains robust even under non-equilibrium dynamics
~\cite{Nunez2025}. 
The perspectives provided by calculating $Q$ are less well explored when it comes to bulk-interacting 
systems, although the covariance matrix has been used as a diagnostics tool for many-body 
localization ~\cite{Bera2015} or in quench protocols~\cite{Thamm2022}.
It was also pointed out recently that an entropy characterizing many-body complexity can be defined based 
on the covariance matrix spectrum~\cite{Vanhala2024}. 
Here we complement such general perspectives by focusing on correlated insulators. We show that the spectrum 
of the covariance matrix of a bulk interacting system unveils essential information enabling
the construction of an accurate and tractable approximation for the many-body ground state that
becomes asymptotically exact at strong interactions.

The framework for our study is the $t$-$V$ model in one dimension~\cite{Cloizeaux1966}. 
It comprises spinless electrons with nearest neighbour hopping $t$ and repulsion $V$. It is arguably the simplest model 
to host a quantum phase transition from a Luttinger liquid to a correlated insulator at finite repulsive interactions. 
The $t$-$V$ model is Bethe ansatz solvable~\cite{Yang1966}, and also amenable to controlled numerical solutions via the 
Density Matrix Renormalization Group (DMRG)~\cite{White1992,Karrasch2012}. 
Our primary aim is to understand how features of the covariance matrix translate into the
microscopic structure of electronic correlations, thereby revealing regimes of emergent simplicity, which 
remained obscure in the representations of the state employed in the Bethe ansatz or DMRG.
While the reduced density matrix (and the more general spin-spin correlation functions) of the XXZ-model
have been computed from integrability~\cite{Jimbo1992,Gohmann2005,Kitanine2000,Babenko2021}, 
fermionic counterparts in the $t$-$V$ model defy analytical calculation due to the nonlocal nature of the mapping between spins
and fermions.

Recently Gebhard et al.~\cite{Gebhard2022} implemented a self-consistent perturbative expansion for the $t$-$V$ model on top of the
Hartree-Fock 
approximation,
that remarkably captures the quantum phase transition at finite repulsion, something that Hartree-Fock
theory on its own
fails at. However, this treatment yields inaccurate results for the covariance matrix
spectrum at large repulsion $V \gg t$. The question of the asymptotically correct post-Hartree-Fock treatment 
in this regime is therefore still open. We show that the solution can essentially be read off from the
perturbatively calculated correlation matrix spectrum. The key is to employ Wannierized natural
orbitals to construct a state whose variational energy agrees with strong-coupling perturbation
theory in next-to-next-to leading order, one order higher in $(t/V)^2$ than Hartree-Fock theory. 
In the Wannierized natural orbital basis, this correlated state consists of a superposition of terms in which the
Hartree-Fock state is decorated by an arbitrary number of identical length four disruptions of the
mean-field order (namely, the charge pattern is flipped along four consecutive
sites). The amplitude of each of the exponentially many terms is exponentially small in 
the number of disruptions. As a result, the post-Hartree-Fock ground state of the $t$-$V$ model 
is a particularly simple example of a class of states known in quantum chemistry as Coupled Cluster
states~\cite{Bartlett2007}. Connecting the many-body states of correlated insulators, a general topic 
of condensed matter, to one of the core concepts of quantum chemistry is our first major result. 
It should be noted that Coupled Cluster methods have already been brought to bear on
extended systems \cite{Hirata2004}, including applications in Condensed Matter \cite{Bishop1978,Shepherd2012,Georgio2024}.
In these studies, Coupled Cluster states are invoked as part of a versatile but involved {\em ab initio} method,
in conjunction with application-specific approximations, similar in spirit to diagramatic Green function methods,
to obtain approximate results that are however not variational. This stands in contrast to the simple and variational
Coupled Cluster picture that we uncover in the $t$-$V$ model at large $V$.

As a next decisive step, we explicitly construct a Matrix Product state (MPS) representation~\cite{Cirac2021} 
of the Coupled Cluster state. The bond dimension of the Matrix Product state is exactly four,
and corresponds precisely to the size of the individual disruptions in the Coupled Cluster state. 
As far as we know, Matrix Product and Coupled Cluster states (CCS) have enjoyed only limited synergy 
\cite{Sharma2015,Veis2016} prior to our work, and the exact mapping between two instances of CCS 
and MPS constitutes our second major result. 

A third important result is provided by an explicit connection between the Coupled Cluster state and 
the statistical mechanics of a one-dimensional lattice gas of tetramers (namely, rods of length
four)~\cite{Ramirez-Pastor1999}. Using a quantum-classical mapping, we derive a novel connection between 
the strong interaction regime of the $t$-$V$ model and solvable quantum tetramer Hamiltonians of the 
Rokhsar-Kivelson form~\cite{Castelnovo2005}, initially introduced for dimers in the context of 
resonating valence bond states~\cite{Rokhsar1988}.
Our work clearly establishes several previously hidden connections between different non-perturbative 
methods for interacting fermions: Coupled Cluster states, Matrix Product states, and
Rokhsar-Kivelson states.

Both the connection to Coupled Cluster and Rokhsar-Kivelson states allow us to derive parent
Hamiltonians for which the Coupled Cluster state that we identified is the exact ground state. 
These two parent Hamiltonians are of interest in their own right, and have not been previously
investigated as far as we know.
(Note: a parent construction for the low-rank Matrix Product state is also possible, but is already
well established~\cite{Fannes1992,Fernandez-Gonzalez2015}, and will not be considered in detail).
The existence of exact parent Hamiltonians invites
the study of the Coupled Cluster state at arbitrary correlation 
strengths, leading us to investigate new physical regimes that are inaccessible within the original 
$t$-$V$ model. We obtain here an exact and compact expression for the covariance matrix elements of 
our Coupled Cluster state in the thermodynamic limit and at arbitrary correlation strength. 

The rest of this Article is structured as follows. In Sec.~\ref{sec:model}, we introduce the
one-dimensional $t$-$V$ model and set notation. In Sec.~\ref{sec:ungapped}, we briefly consider the
critical point for the case of attractive interactions. In this case, we calculate the covariance matrix exactly,
and provide a closed form expression (\ref{eq:Qexact}) for its matrix elements at finite size, which
we could not find in the existing literature. When we consider large repulsive interactions in
subsequent sections, we will discover an interesting parallel with the critical state considered
here. In Sec.~\ref{sec:pert}, we examine the strong-coupling perturbation expansion of the
covariance matrix of the $t$-$V$ model, and notice a striking feature which determines the correct
post-Hartree-Fock treatment in this regime. In Sec.~\ref{sec:cc} we translate the insights we
gained in the previous section into a Coupled Cluster ansatz. We show how the calculation of
physical observables amount to a combinatorial problem in classical statistical physics. 
We numerically demonstrate the accuracy of the Coupled Cluster state by comparing to Bethe ansatz 
and DMRG results for the ground state energy and covariance matrix spectrum. In Sec.~\ref{sec:mps}, 
we analytically construct an exact Matrix Product State representation for the Coupled Cluster state. 
In Sec.~\ref{sec:parent}, we derive two parent Hamiltonians for which the Coupled Cluster state is the unique ground state.
Viewing the Coupled Cluster state as the ground state of either Hamiltonian, gives us license to
consider arbitrary correlation strengths. In Sec.~\ref{sec:lalpha} we present an elegant derivation
of an exact expression for the non-zero matrix elements of the covariance matrix associated with the
Coupled Cluster state at arbitrary correlation strength, before summarizing our conclusions and
giving perspectives in Sec.~\ref{sec:conclusions}. Various technical details can be found in Appendices. 
 
\section{Model}
\label{sec:model}
We consider here the $t$-$V$ model of spinless fermions in one dimension~\cite{Cloizeaux1966}
\begin{align}
H=&-t\sum_{j=1}^{2N}\left(a_j^\dagger a_{j+1}+a_{j+1}^\dagger a_{j}\right)\nonumber\\
&+V\sum_{j=1}^{2N}\left(a_j^\dagger a_{j}-1/2\right)\left(a_{j+1}^\dagger a_{j+1}-1/2\right).\label{eq:tvham}
\end{align}
The $t$-$V$ model is fully equivalent to the solvable one-dimensional
anisotropic Heisenberg (or XXZ) model~\cite{Yang1966} using Jordan-Wigner
transformation~\cite{Giamarchi2003}.
Without loss of generality, we can take $t>0$.
Periodic boundary conditions are implied, i.e. $a_j=a_{j+2N}$, unless noted otherwise
(in particular for the construction of the MPS representation).
The $t$-$V$ Hamiltonian is particle-hole symmetric, i.e. $H=PHP$ where 
\begin{equation}
P=\prod_{j=1}^N \left(a_{2j-1}-a_{2j-1}^\dagger\right)\left(a_{2j}+a_{2j}^\dagger\right)\label{eq:phcon}
\end{equation} 
is a Hermitian and unitary particle conjugation operator such that $Pa_jP=(-1)^ja_j^\dagger$.
We will always work at half-filling.
It is known that the Hamiltonian (\ref{eq:tvham}) undergoes quantum phase transitions at $V=\pm2t$.
For $V>2t$ there are two macroscopically distinct gapped ground states, that each represent a correlated insulator,
with charge order, associated to broken particle-hole conjugation,
as well as broken symmetry between the two sublattices of the original bipartite lattice.
Because our main focus is this correlated insulator phase, we will work throughout with two-site 
unit cells $(2j-1,2j)$ and used folded Bloch momenta that are rescaled in the range $k\in(-\pi,\pi]$, corresponding 
to the range $(-\pi/2,\pi/2]$ of the unfolded first Brillouin zone in the original lattice.

\begin{center}
\begin{figure}
\includegraphics[width=0.99\columnwidth]{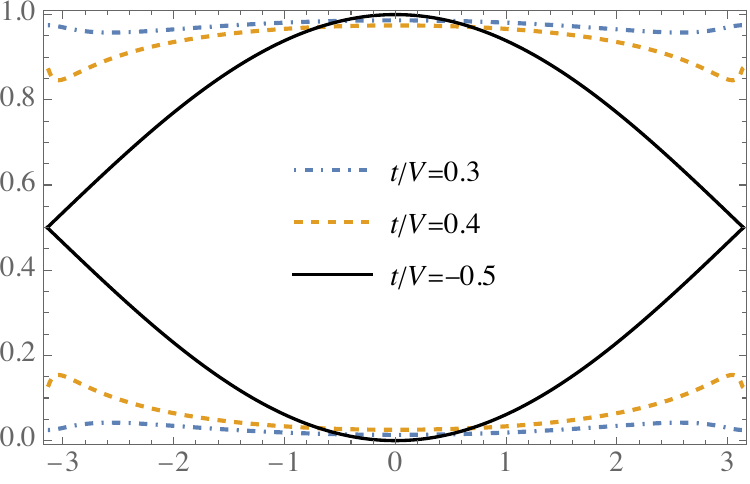}
\caption{\label{fig:qmatspec} The behavior of the covariance matrix spectrum in the rescaled and folded Brillouin zone. 
The dashed lines are numerical DMRG data for the gapped correlated insulator 
phase $V>2t$, while the dashed curve is the analytical formula Eq.~(\ref{eq:qkcrit}) for the critical point $V=-2t$ 
associated to a uniform gapless state.} 
\end{figure}
\end{center}

In the Luttinger liquid phase, the unbroken sub-lattice symmetry implies that the natural orbitals are simply the single-particle momentum eigenbasis.
All physical information contained in the covariance matrix then resides in its spectrum, which corresponds to their occupation numbers $q_{\pm}(k)$.
The subscript $\pm$ is necessary because of the folding of the Brillouin zone, i.e. $q_-$ ($q_+$) refers
to occupations of orbitals with momenta outside (inside) the interval $(-\pi/2,\pi/2]$ in the unfolded Brillouin zone. 
For $V=0$, the ground state is a degenerate Fermi gas with $q_-(k)=0$ and $q_+(k)=1$ for all $k$.
In the rest of the ungapped phase, namely for $|V|\leq 2t$, it is known~\cite{Giamarchi2003} 
that $q_+(k)$ and $q_-(k)$ touch at $k=\pm \pi$, where particle hole 
symmetry protects $q_\pm(k)=1/2$, reflecting the fact that arbitrarily weak interactions destroy Fermi liquid behavior in one dimension. 
At $V=\pm 2 t$, where lattice translation symmetry is spontaneously broken, gaps start to appear at $k=\pm\pi$.
In Figure~\ref{fig:qmatspec}, the behavior of the covariance matrix spectrum in the correlated
insulator phase is shown. Since the ground state in the $V\to\infty$
limit is an uncorrelated single Slater determinant, the eigenvalues of the covariance matrix move
closer to $0$ or $1$, as $t/V$ decreases towards zero.
The dashed lines are numerically exact DMRG data, obtained using the C++ ITensor library~\cite{ITensor2022} 
for long chains with $2N=2000$ sites, using up to 30 sweeps, a maximum bond dimension $D=300$, and a truncation 
threshold $10^{-14}$.
The covariance matrix spectrum of the translationally invariant system was obtained
from a long chain with open boundaries (which helps convergence of DMRG) by computing
the $Q$-matrix 
$Q_{jj'} =\big<a_j^\dagger a_{j'}\big>$ for sites $j,j'$ near the center of the chain, in a region 
where translation invariance was carefully checked.
In addition, the solid black curve in Fig.~\ref{fig:qmatspec} indicates the analytically obtained gapless 
spectrum at $V=-2t$. (See Sec.~\ref{sec:ungapped} below.)
It shows the touching points at $k=\pm\pi$ that are generic for the whole gapless phase.

\section{The critical point at $\mathbf{V=-2t}$}
\label{sec:ungapped}

Before studying the correlated insulator phase, we briefly focus on the gapless critical point 
at $V=-2t$, which is exactly solvable, and provides already some useful insights.
In subsequent sections, we uncover interesting parallels and differences between this gapless ground state
and the gapped insulating state found at large positive $V$. 

At $V=-2t$, the critical ground state is the equal weight superposition of all arrangements of $N$
particles among $2N$ sites, see~\cite{Barghathi2019} or Appendix \ref{app:GS_crit}, and thus reads:
\begin{equation}
\left|\text{GS}\right>_{V=-2t} = {2N \choose N}^{-1/2}\sum_{\gamma\in C(2N,N)}
a_{\gamma_1}^\dagger a_{\gamma_2}^\dagger\ldots a_{\gamma_N}^\dagger\left|0\right>.\label{eq:GS_crit}
\end{equation}
Here each $\gamma=(\gamma_1,\gamma_2,\ldots,\gamma_N)$ consists of $N$ lattice indices chosen from the
$2N$ sites of the chain, arranged such that $\gamma_1<\gamma_2<\ldots<\gamma_N$, and the sum runs
over all possible choices. For this state, we obtain analytically in Appendix~\ref{app:crit} the 
matrix elements of the covariance matrix in position representation, $Q_{j,j}=1/2$, while for $m>0$,
\begin{equation}
Q_{j,j+m}= \frac{\sum_{n=\max(0,m-N)}^{\min(m-1,N-1)}(-1)^n{m-1 \choose n}{2N-m-1 \choose N-n-1}}{{2N \choose N}}.\label{eq:Qcrit}
\end{equation} 
This expression is easy to grasp. When $a_1^\dagger a_{1+m}$ is applied to a term in the sum
(\ref{eq:GS_crit}), it yields a non-zero result if site $1$ is empty and site $1+m$ is occupied.
Each of these states is mapped into another state in the sum, up to a sign, and thus contributes
$\pm{2N \choose N}^{-1}$ to the overlap with the ground state. The sign is determined by the number
$n$ of particles between sites $1$ and $1+m$. We count the number of terms with site $1$ empty and
site $1+m$ occupied, that have $n$ particles on sites $2$ to $m$ as follows. There are ${m-1 \choose
n}$ ways to arrange the $n$ particles on sites $2$ to $m$, and ${2N-m-1 \choose N-n-1}$ ways of
arranging the remaining $N-n-1$ particles among the remaining $2N-m-1$ sites. By translation
invariance $Q_{j,j+m}=Q_{1,1+m}$. The sum in (\ref{eq:Qcrit}) can be evaluated analytically. It is
zero for $m$ even. (This is guaranteed by particle-hole symmetry.) For $m=2r+1\leq N$ 
\begin{equation}
Q_{j,j+2r+1}=(-1)^r\frac{2^{2(N-1)}\Gamma(r+1/2)\Gamma(N-r-1/2)}{(N-1)!\Gamma(1/2)^2 {2N \choose N}}.\label{eq:Qexact}
\end{equation}
(For $m>N$ we have $Q_{j,j+m}=Q_{j,2N+1-m}$ thanks to inversion symmetry.)
It can be verified that $Q_{j,j+2r+1}$ decays like $N^{-r}$.
See Appendix~\ref{app:crit} for details on the derivation.
To the best of our knowledge, the closed form expression (\ref{eq:Qexact}) does not appear in the
literature, although it is known that the correlation length of the covariance matrix vanishes in the thermodynamic limit at
$V=-2t$~\cite{Thamm2022}. The extremely rapid decay of off-diagonal elements of the covariance
matrix is a destructive interference effect between the contribution by states with an even or an
odd number of fermions on sites $j+1,\ldots,j+2r+1$. This fermionic sign effect was pointed out in
\cite{Kohno2010}. In the thermodynamic limit 
\begin{equation}
Q_{j,j'}=\frac{1}{4}\left(2\delta_{j,j'}+ \delta_{j,j'+1}+\delta_{j+1,j'}\right).\label{eq:Qthermo}
\end{equation}
Choosing to work in the folded Brillouin zone, we diagonalize $Q$ by means of the unitary transform $Q'=F Q F^\dagger$ where 
\begin{eqnarray}
\left(\begin{array}{ll} F_{2j-1,2l-1} & F_{2j-1,2l}\\
F_{2j,2l-1}&F_{2j,2l}\end{array}\right)&=&\frac{e^{i k_l j}}{\sqrt{N}}\left(\begin{array}{cc} -\frac{e^{-ik_l/2}}{\sqrt{2}}&\frac{1}{\sqrt{2}}\\
\frac{1}{\sqrt{2}}&\frac{e^{ik_l/2}}{\sqrt{2}}\end{array}\right),\nonumber\\
\left(\begin{array}{ll} Q'_{2l-1,2l'-1} & Q'_{2l-1,2l'}\\
Q'_{2l,2l'-1}&Q'_{2l,2l'}\end{array}\right)&=&\delta_{l,l'}\left(\begin{array}{cc}q_-(k_l)&0\\0&q_+(k_l)\end{array}\right),
\end{eqnarray}
for $k_l=2\pi \left(l-\frac{N-1}{2}\right)/N$, $l=1,2,\ldots,N$, and the spectrum of the covariance matrix is simply
\begin{equation}
q_\pm( k)=[1\pm \cos(k/2)]/2.\label{eq:qkcrit}
\end{equation} 
Due to our choice of working in a folded Brillouin zone, 
the $Q$-matrix spectrum comprises two bands.
These bands have $4\pi$ periodicity, due to the full lattice symmetry of the ground state,
and touch at $k=\pm\pi$, as expected in the gapless regime. Let us now contrast these results
with 
the spectrum obtained for large electron-electron repulsion $V/t\gg1$.

\section{Large $\mathbf{V/t}$ perturbation theory}
\label{sec:pert}

From here onwards, we focus on the correlated insulator phase $V>2t$. Our starting point is to
consider the spectrum of the covariance matrix at strong repulsion. In the small $t/V$ limit,
Rayleigh-Schr\"odinger perturbation theory around the unperturbed ground state 
\begin{equation}
\left|\Psi_0\right>=a_2^\dagger a_4^\dagger\ldots a_{2N}^\dagger\left|0\right>,\label{eq:psi_0}
\end{equation}
yields the following result for the
spectrum of the covariance matrix, to fourth order in $t/V$:
\begin{equation}
q(k)_\pm=\frac{1}{2}\pm\left[\frac{1}{2}-(4-2\cos 2k)\left(\frac{t}{V}\right)^4\right].\label{eq:qk}
\end{equation} 
See Appendix~\ref{app:covmat} for details:
remarkably, the spectrum has $k$-periodicity $\pi$, even though the matrix $Q(k)$ only has $2\pi$ $k$-periodicity.
What does this quartering of the period compared to (\ref{eq:qkcrit}) tell us about the structure of the gapped ground state? 

The formal perturbative expansion of the ground state wave function only answers this question in an indirect way.
In the present context, this general statement can be elaborated as follows.
When the perturbation $H_{\text{kin}}=-t\sum_{j=1}^{2N}\left(a_j^\dagger a_{j+1}+a_{j+1}^\dagger a_{j}\right)$ acts
on the unperturbed state (\ref{eq:psi_0}), it moves a single particle from the $B$-sublattice (even sites) to the $A$-sublattice.
Thus the expansion of the wave-function to $n$-th order contains at most $n$ particles on the $A$ sublattice.
However, the true ground state at any non-zero $t/V$ hosts on average a finite fraction of the total number of particles on the $A$-sublattice.
In the thermodynamic limit, the formal perturbation expansion to finite order thus poorly approximates the true ground state.
A symptom of this state of affairs is that the perturbatively calculated
ground state energy is not variational. In fact, if we denote the ground state wave function and energy, expanded to $n$th order
as respectively $\left|\Psi^{(n)}\right>$ and $E_0^{(n)}$, then due to the non-extensive nature of ground state corrections 
\begin{equation}
\lim_{N\to \infty} \frac{1}{N} \frac{\left<\Psi^{(n)}\right|H\left|\Psi^{(n)}\right>}{\left<\Psi^{(n)}\right|\left.\Psi^{(n)}\right>}
= E_0^{(0)},\label{eq:pertvar}
\end{equation}
i.e. in the thermodynamic limit, finite order perturbation theory is variationally no better than
the unperturbed starting point. To obtain the answer $E_0^{(n)}$ correct to order $n$ 
instead
of $E_0^{(0)}$, the above expression must be expanded to order $n$ before the limit $N\to\infty$ is
taken. An order-by-order cancellation of certain terms between numerator and denominator then
occurs. In the diagramatic Green's function formalism, this corresponds to the cancellation of
disconnected diagrams, leaving as final result the sum of connected diagrams. This allows us to
calculate expectation values of simple observables. However, the unsatisfactory
fact remains that one cannot explicitly write a thermodynamically consistent perturbative wave function, 
and ask simple questions, such as how accurately can the perturbative ground state be approximated for 
a given MPS bond dimension?

More direct insight than that offered by perturbation theory of the ground state wave function
would be provided by a state whose variational energy equals $E_0^{(n)}+\mathcal O (t/V)^{n+1}$ for some $n$.
Standard mean-field theory solves this problem to first non-vanishing order in the perturbation,
namely at order $(t/V)^2$.
Here we pursue the quest for a state that is accurate to the next non-vanishing order, 
namely at order $(t/V)^4$. This is the order necessary to shed light on the emergent
$\pi$-periodicity of the covariance matrix spectrum, since the correlation spectrum of
the mean-field state is trivial (eigenvalues are either 0 or 1). Our main task is thus
to construct a state describing a finite density of quantum fluctuations of the mean-field order
parameter, with a variational energy that agrees with the perturbative result
\begin{equation}
E_0=-\frac{NV}{2}\left[1+4\left(\frac{t}{V}\right)^2-4\left(\frac{t}{V}\right)^4+\ldots\right],\label{eq:E_0}
\end{equation}
to order $\left(\frac{t}{V}\right)^4$.
[The expansion (\ref{eq:E_0}) can be verified from the exact Bethe ansatz expression (\ref{eq:exact}) below.]

\section{Coupled Cluster doubles ansatz}
\label{sec:cc}

\begin{center}
\begin{figure}
\includegraphics[width=0.99\columnwidth]{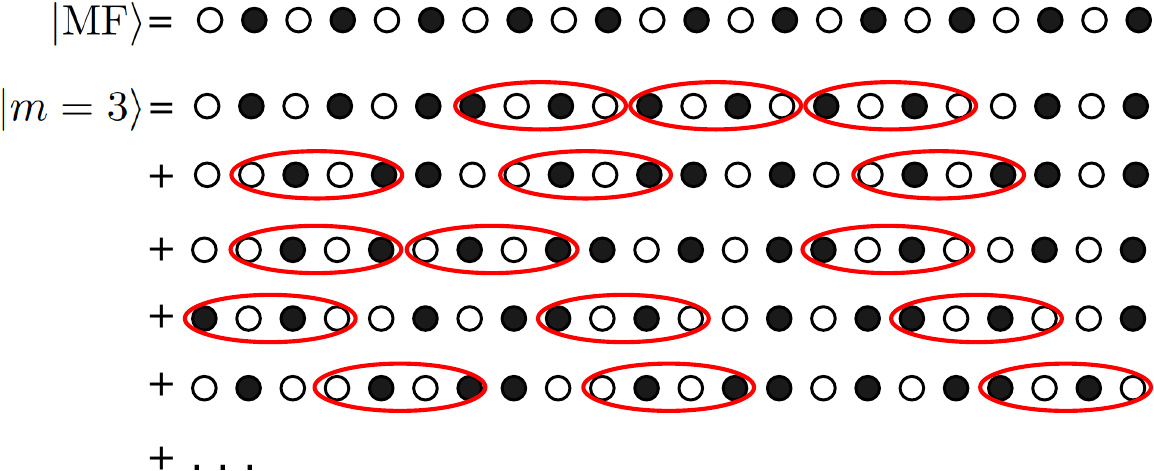}
\caption{\label{fig:state} Schematic representation of the mean-field state
$\left|\text{MF}\right>$ and the state $\left|m=3\right>$ built from three disruptions of length
four (red ellipses) on top of the mean-field state, for $N=11$ fermions in a chain of $2N=22$ sites. 
Filled/empty circles represent the center of mass of filled/empty Wannierized Hartree-Fock
orbitals (and not localized sites of the original lattice).
For $\left|m=3\right>$, five of the $484$ configurations in the equal weight superposition (\ref{eq:m}) are
depicted here.} 
\end{figure}
\end{center}

The Hartree-Fock mean-field theory of the correlated insulator at small $t/V$ is reviewed in
Appendix \ref{app:mf}. It yields a set of new fermion operators $c_j$, $j=1,\ldots,2N$, with the
following features. The Wannierized Hartree-Fock basis is invariant under translations by two lattice sites, 
under which action $c_j\to c_{j+2}$. The single-particle orbital associated with $c_j$ is peaked at site $j$,
and has a more pronounced amplitude on sites with the same parity as $j$, than on sites of
opposite parity. 

In this basis of Wannierized Hartree-Fock orbitals, the mean-field state reads 
\begin{equation}
\left|\text{MF}\right>=c_2^\dagger c_4^\dagger\ldots c_{2N}^\dagger \left|0\right>.\label{eq:mfvec}
\end{equation}
Of course, the same state could equivalently be obtained as the usual Slater determinant 
in which all Bloch momenta of the lower band of the mean field Hamiltonian are occupied,
but the Wannierized version is more 
amenable to constructing a correlated state that is energetically
favorable, since the repulsive term is quasi-local in space, as we discuss now.

Our ansatz introduces order parameter fluctuations on top of 
the state (\ref{eq:mfvec}).
These take the form of regions of opposite order (even sites empty, odd sites filled) with respect to the
mean-field pattern~\cite{Fratini2004,Mayr2006}.
Our key insight for reproducing the $\pi$-periodicity of the spectrum of the covariance matrix
is that the fluctuations should be made of building blocks of size two unit cells of the symmetry broken system, 
i.e. four sites of the original lattice. We call four adjacent sites of which the two even orbitals $c_{2j}$ and $c_{2j+2}$
are empty, while the two odd orbitals $c_{2j+1}$ and $c_{2j+3}$ are filled, a disruption, depicted with
a red ellips in Figure~\ref{fig:state}. Sites with the same occupation pattern as 
$\left|\text{MF}\right>$ -- odd sites empty, even ones filled -- are said to be of regular order.
Again, we point out that the $c_j^\dagger$ fermions on ``sites'' $j$ correspond to Wannierized orbitals 
centered around the real space position $j$.
The four-fermion operator
\begin{equation}
T=\sum_{j=1}^{N} \left( c_{2j} c_{2j+1}^\dagger c_{2j+2} c_{2j+3}^\dagger 
+c_{2j-1}^\dagger c_{2j}c_{2j+1}^\dagger c_{2j+2}\right),\label{eq:T}
\end{equation}
creates one such a disruption at any of these possible locations ($l-3$, $l-2$, $l-1$, $l$) of the lattice, 
in a translationally invariant manner. The unnormalized state
\begin{equation}
\left|m\right>=\frac{1}{m!}T^m\left|\text{MF}\right>,\label{eq:m}
\end{equation}
is a superposition of all possible ways to make $m$ disruptions, in which each unique term appears
with weight equal to unity. (See Figure~\ref{fig:state}.) Note that the product of any two terms in $T$ that
are associated with overlapping disruptions is zero. Disruptions therefore correspond to hard-core rods of size four. 
The norm of $\left|m\right>$ is obtained by counting the
number of non-zero terms in $T^m$. There are $m$ disruptions and $2N-4m$ sites of regular order to arrange into a
string. This gives a combinatorical factor ${2N-3m \choose m}$. This way of counting includes a
factor $2N-3m$ corresponding to the number of cyclic permutations of the string, in which a
disruption is never allowed to straddle the ``seam'' where site $2N$ is glued to site $1$. We must
however also include the arrangements where a disruption straddles the $2N$ --- $1$ seam, and
therefore must multiply by $2N/(2N-3m)$, i.e.
\begin{equation}
\left<m\right|\left.m\right>=\frac{2N}{2N-3m}{ 2N-3m \choose m}.\label{eq:norm}
\end{equation}

If we assume that the fluctuations caused by multiple disruptions are independent of each other and
are each associated with an amplitude $\alpha$, then the resulting state is known as a coupled
clusters doubles state built from the operator $T$~\cite{Bartlett2007}. It reads:
\begin{equation}
\left|\alpha\right>=\exp(\alpha T)\left|\text{MF}\right>=\sum_{m=0}^{\lfloor N/2\rfloor}\alpha^m\left|m\right>.\label{eq:ansatz}
\end{equation}
(Note that $\left|\alpha\right>$ is not normalized.) 
The exponential term $\exp(\alpha T)$ takes care of size extensivity by inducing a similarity transform
$\exp(\alpha T) c_{2n}^\dagger \exp(-\alpha T)$ on each creation operator in (\ref{eq:mfvec}). Each
transformed operator has a finite amplitude to create particle-hole excitations, and as a result the
typical number of total excitations is proportional to $N$. See (\ref{eq:simtrans1}) and
(\ref{eq:simtrans2}) below.
 
Usually, the extensive nature of Coupled Cluster states makes calculations for more than a
handful of particles impossible, due to the exponential number of involved terms. We emphasize
that this statement applies to calculations involving the full Coupled Cluster wave function, and not 
to the linearized treatment of Coupled Cluster equations, that has an algebraic complexity, but is 
not variational.
Remarkably, the simple structure of the disruptions in the variational Coupled Cluster state that we consider
avoids paying an exponential price. Indeed, although the states $\left|m\right>$ are many-body correlated 
states, for which Wick's theorem does not apply, operator matrix elements can nonetheless be calculated by
combinatorial means. 

We now sketch how analytical computations of the variational energy and the $Q$-matrix can be done 
with the disrupted states~(\ref{eq:m}).
When a single-particle additive operator $c_j^\dagger c_{j'}$ acts on a term in $\left|m\right>$
such as one of those depicted in Figure~\ref{fig:state}, it can only do one of the following things: (i)
leave the term unchanged (if and only if $j=j'$), (ii) annihilate the term, (iii) move a sequence of
head-to-toe disruptions by one site, thereby producing another term in $\left|m\right>$ (up to a
sign), or (iv) produce a state orthogonal to all $\left|m\right>$. Since $c_j^\dagger c_{j'}$ cannot
create or annihilate a full disruption, $\left<m\right|c_j^\dagger c_{j'}\left|m'\right>=0$ if
$m\not =m'$. Expectation values of single-particle additive operators can be calculated by solving
counting problems similar to the one we solved to compute the norm. For instance, $c_{2l}^\dagger
c_{2l}$ counts terms that have a site of regular order at site $2l$ while $c_{2l+1}^\dagger
c_{2l+1}$ counts terms with a disruption involving site $2l+1$. This straightforwardly yields 
\begin{eqnarray}
\left<m\right|c_{2l}^\dagger c_{2l}\left|m\right>={2N-3m-1 \choose m},\nonumber\\
\left<m\right|c_{2l+1}^\dagger c_{2l+1}\left|m\right>=4 {2N-3m-1 \choose m-1}.\label{eq:diag}
\end{eqnarray}
Process (iii) can occur when $c_1^\dagger c_{1+4l}$ moves $l$ head-to-toe disruptions initially
occupying sites $2,\ldots, 4l+1$, one site to the left. An example is illustrated below.\\
\centerline{\texttt{...0[0101]...} $\to$ \texttt{...[1010]0...}} A single four-site disruption is
depicted for simplicity, although we consider the general case of multiple head-to-toe disruptions.
Here \texttt{[...]} indicates the disruption that is being moved. Moving the $l$ disruptions requires
commuting $c_{1+4l}$ past $2l-1$ $c^\dagger$ operators, which is an odd number. If site $1$ is
initially empty, and sites $2,\ldots,4l+1$ are occupied by $l$ disruptions, then we have $2N-4m-1$
remaining sites of regular order and $m-l$ disruptions to arrange in a string. Hence 
\begin{equation}
\left<m\right|c_1^\dagger c_{1+4l}\left|m\right>=-{2N-3m-l-1 \choose m-l},\label{eq:twoc}
\end{equation} 
for $0< l\leq m$.
For $c_2^\dagger c_{2+4l}$ essentially the same considerations apply. The only difference is that
moving $l$ disruptions e.g. \centerline{\texttt{...[0101]1...} $\to$ \texttt{...1[1010]...}} now
requires commuting $c_{2+4l}$ past two $c^\dagger$ operators per disruption, and so no minus sign is
generated. Including also processes that straddle the link between sites $2N$ and $1$ we find 
\begin{align}
&\left<m\right|c_2^\dagger c_{2+4l}\left|m\right>=\left<m\right|c_2^\dagger c_{2N-4l+2}\left|m\right>\nonumber\\
=&-\left<m\right|c_1^\dagger c_{1+4l}\left|m\right>=-\left<m\right|c_1^\dagger c_{2N-4l+1}\left|m\right>.
\end{align}
All non-zero $c_j^\dagger c_{j'}$ expectation values can be obtained from the above by invariance
under translation by two lattice sites, or by Hermitian conjugation. 

We note that every $4l$'th diagonal (modulo $2N$) above or below the main diagonal of 
$Q_{jj'}=\left<\alpha\right|c_j^\dagger c_{j'}\left|\alpha\right>/\left<\alpha\right|\left.\alpha\right>$ 
is non-nonzero, and of order $\alpha^{2l}$ , with all other diagonals above or below the main
diagonal equal to zero. This results in the $\cos(2k)$ component of the covariance matrix spectrum
(coming from the $\pm 4$'th diagonals) being of order
$\alpha^2$. The covariance matrix spectrum expanded to order $\alpha^2$, calculated from
(\ref{eq:diag}) and (\ref{eq:twoc}),
is given by $q_\pm(k)=1/2 \pm\left[1/2-\alpha^2(4-2\cos 2k)\right]$. We thus see that we have to choose 
\begin{equation}
\alpha=\left(\frac{t}{V}\right)^2,
\end{equation}
to ensure agreement with the desired covariance matrix (\ref{eq:qk}) for the ground state. 

Apart from the processes already discussed, two-particle additive operators $c_{j_1}^\dagger c_{j_2}
c_{j_3}^\dagger c_{j_4}$ produce further processes: (v) two strings of head-to-toe disruptions can
each be moved independently, e.g. 
\begin{align}
&\texttt{... 0[0 1 0 1]...~0[0 1 0 1]...}\to\nonumber\\
&\texttt{...[1 0 1 0]0~...[1 0 1 0]0~...};\nonumber
\end{align}
(vi) three strings of head-to-toe disruptions (of which some may be zero strings) can be fused into
a single string containing one more disruption, e.g. 
\begin{align}
&\texttt{... 0[0 1 0 1]1[1 0 1 0]0[0 1 0 1]1...}\to\nonumber\\
&\texttt{...[1 0 1 0 1 0 1 0 1 0 1 0 1 0 1 0]...};\nonumber
\end{align}
(vii) and the reverse of process (vi) in which the number of disruptions decreases by one. We work
out the matrix elements for two-particle additive operators, by solving the appropriate counting
problems in Appendix \ref{app:matels}. 

\begin{center}
\begin{figure}
\includegraphics[width=0.99\columnwidth]{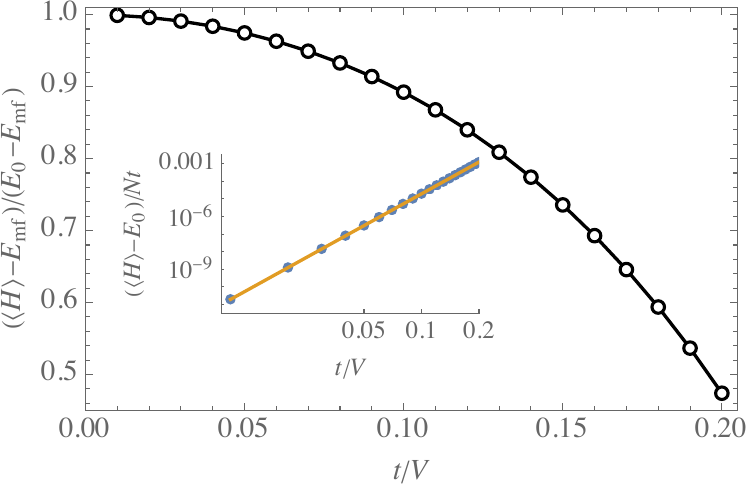}
\caption{\label{fig:energy} {\bf Main panel:} Difference in ground state energy
between mean-field theory (\ref{eq:mfenergy}) and the correlated ansatz (\ref{eq:ansatz}) in units of
the exact correlation energy of the $t$-$V$ model. {\bf Inset:} The difference between the variational 
energy per particle of the correlated ansatz (\ref{eq:ansatz}) and the true ground state energy (\ref{eq:exact}) 
in double log scale. The solid line corresponds to $19.5 (t/V)^6$, and indicates that the variational
energy is correct to order $(t/V)^4$. Results are shown for a system of $N=41$ particles
occupying $82$ sites.} 
\end{figure}
\end{center}

In this way we obtain a variational estimate
$\left<H\right>=\left<\alpha\right|H\left|\alpha\right>/\left<\alpha\right|\left.\alpha\right>$ with
respect to the state $\left|\alpha=(t/V)^2\right>$ of the ground state energy, that includes
non-trivial correlation effects. See Appendix~\ref{app:ham} for further details. To investigate the
quality of the correlated ansatz, we consider the correlation energy $E_\text{mf}-\big< H \big>$, 
where $E_\text{mf}$ (\ref{eq:mfenergy}) is the best estimate of the ground state energy that can be
obtained from an uncorrelated (mean-field) trial state. 
The exact ground state energy is known via the Bethe ansatz~\cite{Yang1966}.
\begin{equation}
\frac{E_0}{N}=t\left[\cosh^{-1}\lambda -2\sinh^{-1}\lambda\left(1+4\sum_{n=1}^\infty\frac{1}{1+e^{2n\lambda}}\right)\right],\label{eq:exact}
\end{equation}
where $\lambda=\cosh(V/2t)$. The main panel of Figure~\ref{fig:energy} shows
the correlation energy $(E_\text{mf}-\left<H\right>)/(E_\text{mf}-E_0)$ relative to the exact value, 
for increasing $t/V$.
At small $t/V$ the ratio tends to unity, implying that in this regime, the correlated ansatz accounts for the full effect of correlations. 
The inset of Figure~\ref{fig:energy} shows that the difference $\left<H\right>-E_0$ grows like $(t/V)^6$ at small $t/V$. This indicates that
the correlated ansatz gives a variational energy accurate to fourth order $\sim(t/V)^4$. In
contrast, Hartree-Fock mean-field theory gives a variational energy that is only accurate to 
order $\sim(t/V)^2$. (See Appendix~\ref{app:mf}.) The correlated ansatz (\ref{eq:ansatz}) thus
achieves the goal of building a size-extensive post-Hartree-Fock wavefunction, that we had previously set 
in Sec.~\ref{sec:pert}. 
\begin{center}
\begin{figure}
\includegraphics[width=0.99\columnwidth]{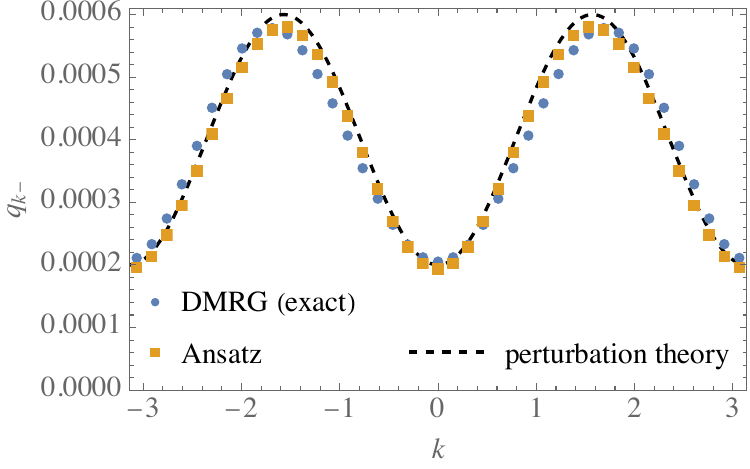}
\caption{\label{fig:q-spec} The lower band $q_{k-}$ of the correlation matrix spectrum of the
$t$-$V$ model, at $t/V=0.1$, for a system of $N=41$ particles occupying $82$ sites. The 
variational results for the Coupled Cluster ansatz (\ref{eq:ansatz}) are compared to numerically 
exact calculations obtained with the Density Matrix Renormalization Group (DMRG), and to the perturbative 
expression (\ref{eq:qk}).}
\end{figure}
\end{center}

In Figure~\ref{fig:q-spec}, we plot the lower band of the covariance matrix spectrum,
calculated from the correlated ansatz (\ref{eq:ansatz}) with the help of (\ref{eq:diag}) and
(\ref{eq:twoc}), and compare to numerically exact results obtained with DMRG (to our knowledge, 
the correlation spectrum is not available from the Bethe ansatz), as well as to the perturbative 
result (\ref{eq:qk}). This computation is done for $t/V=0.1$, and for this not-so-small value
of $t/V$, agreement of the ansatz and DMRG with perturbation theory is still good. Deviations between 
the ansatz and DMRG, due to higher order effects, are just becoming visible. From this comparison, 
we conclude that the regime where four-site disruptions of the mean-field order quantitatively accounts 
for the correlation effects is $t/V\lesssim 0.1$. It may be noted that our variational ansatz is very 
restrictive, containing no free parameters to be optimized (once $\alpha$ was set). In
Appendix~\ref{app:var}, we present numerical evidence that the Coupled Cluster doubles form assumed 
here is indeed optimal for a more general ansatz built out of the states $\left|m\right>$.

Armed with a Coupled Cluster ansatz that is variationally accurate in this regime, we now turn to
investigate the approximate Matrix Product state (MPS) structure of the ground state. It turns out 
that an exact MPS representation of the Coupled Cluster state $\left|\alpha\right>$ can be constructed 
analytically, establishing a bridge from ideas about one-dimensional correlated fermions to concepts in quantum chemistry.

\section{Exact MPS representation of the Coupled Cluster state}
\label{sec:mps}

As is often the case in the context of MPS, we find it convenient to consider open boundary
conditions. We consider a system with $L$ sites. Let $D_j$ be the operator that creates a $4$-site
disruption ending on site $j$, i.e.
\begin{equation}
D_j=\left\{\begin{array}{ll} c_{j-3}c^\dagger_{j-2}c_{j-1}c^\dagger_{j}\mbox{ if }j\mbox{ is odd},\\
c^\dagger_{j-3}c_{j-2}c^\dagger_{j-1}c_{j}\mbox{ if }j\mbox{ is even}.
\end{array}\right.\label{eq:defd}
\end{equation}
We recall that the $c_j^\dagger$ fermions on ``sites'' $j$ are associated to Wannierized orbitals
centered around the real space position $j$.

We define a disruption operator acting on sites $1$ to $l$:
\begin{equation}
T_l=\left\{\begin{array}{ll}0\mbox{ if }l<4,\\
\sum_{j=4}^l D_j\mbox{ if }l\geq 4.\end{array}\right.
\end{equation}
We also find it convenient to introduce the following notation for states. Let
$(c_l^\dagger)^1\equiv c_l^\dagger$ and $(c_l^\dagger)^0\equiv 1$. An arbitrary single Slater
determinant in the Wannierized mean-field basis
is then denoted
\begin{equation}
\left| n_1n_2\ldots n_L\right>\equiv 
\left(c_1^\dagger\right)^{n_1}\left(c_2^\dagger\right)^{n_2}\ldots \left(c_L^\dagger\right)^{n_L}\left|0\right>,
\end{equation}
where $n_j\in\{0,1\}$.
Furthermore, let 
\begin{equation}
r_l=\left\{\begin{array}{ll} 0\mbox{ if }l\mbox{ is odd},\\
1\mbox{ if }j\mbox{ is even},
\end{array}\right.
~~~~
d_l=\left\{\begin{array}{ll} 1\mbox{ if }l\mbox{ is odd},\\
0\mbox{ if }j\mbox{ is even}.
\end{array}\right.
\end{equation}
(The notation $r$ and $d$ here denote ``regular'' and ``disrupted'' sites.) In this notation, the mean-field state reads
\begin{equation}
\left|\text{MF}\right>=\left|r_1r_2\ldots r_L\right>,
\end{equation}
while
\begin{equation}
D_j\left|\text{MF}\right>=\left|r_1\ldots r_{j-4} d_{j-3}\ldots d_j r_{j+1}\ldots r_L\right>.
\end{equation}
Then the Coupled Cluster state that we considered previously is
\begin{equation}
\left|\alpha\right>\equiv\exp(\alpha T_L)\left|r_1r_2\ldots r_L\right>.\label{eq:targetstate}
\end{equation}

At the heart of the MPS expansion is the task of relating states of the same form on systems that differ in size by one site.
In view of this, we note that for $l\geq 4$,
\begin{equation}
T_l=T_{l-1}+D_{l}.
\end{equation}
We further note that $D_l$ kills all terms in $T_{l-1}$ that involve sites $l-3,\ldots,l-1$. Hence
\begin{equation}
D_l\left(T_{l-1}\right)^k=D_l\left(T_{l-4}\right)^k.
\end{equation} 
From this together with $[D_j,D_{j'}]=0$, and $D_j^2=0$ follows that
\begin{equation}
e^{\alpha T_l}=e^{\alpha T_{l-1}}+\alpha D_le^{\alpha T_{l-4}}.\label{eq:exprecur}
\end{equation} 

The next step is to define a set of states that have a desired form on sites left of site $l+1$, and an arbitrary form on sites $l+1,\ldots,L$.
We define 
\begin{align}
&\left|1,n_{l+1}\ldots n_L\right>=e^{\alpha T_l}\left|r_1\ldots r_ln_{l+1}\ldots n_L\right>,\\
&\left|2,n_{l+1}\ldots n_L\right>=e^{\alpha T_{l-1}}\left|r_1\ldots r_{l-1}d_ln_{l+1}\ldots n_L\right>,\nonumber\\
&\left|3,n_{l+1}\ldots n_L\right>=e^{\alpha T_{l-2}}\left|r_1\ldots r_{l-2}d_{l-1}d_ln_{l+1}\ldots n_L\right>,\nonumber\\
&\left|4,n_{l+1}\ldots n_L\right> =
e^{\alpha T_{l-3}}\left|r_1\ldots r_{l-3}d_{l-2} d_{l-1} d_ln_{l+1}\ldots n_L\right>\nonumber.
\end{align}
State $\left|j,n_{l+1}\ldots n_L\right>$ is of the Coupled Cluster form on sites $1$ to $l+1-j$, and has arbitrary occupations on sites $l+1$ to $L$. For $j>1$, it contains $j-1$ disrupted
sites situated from site $l+2-j$ to site $l$. 
For $l=0$ we define $\left|1,n_1\ldots n_L\right>\equiv \left|n_1\ldots n_L\right>$.
Employing (\ref{eq:exprecur}), we find
\begin{align}
&\left|1,n_{l+1}\ldots n_L\right>=e^{\alpha T_l}\left|r_1\ldots r_l n_{l+1}\ldots n_L\right>\nonumber\\
&=\left|1,r_l n_{l+1}\ldots n_L\right>+\alpha\left|4,d_l n_{l+1}\ldots n_L\right>.
\end{align}
Furthermore we trivially have
\begin{equation}
\left|j,n_{l+1}\ldots n_L\right>=\left|j-1,d_ln_{l+1}\ldots n_L\right>,
\end{equation}
for $j>1$.
Thus we can write, with summation of repeated indices (including $\sigma$) implied, 
\begin{equation}
\left|j,n_{l+1}\ldots n_L\right>=A(l)^{(\sigma)}_{jj'}\left|j',\sigma n_{l+1}\ldots n_L\right>,
\end{equation}
where for $l\geq 4$, $A(l)^{(\sigma)}$, $\sigma \in\{0,1\}$ is given by
\begin{equation}
A(l)^{(r_l)}=\left(\begin{array}{cccc}1&0&0&0\\
0&0&0&0\\
0&0&0&0\\
0&0&0&0\end{array}
\right),\label{eq:ar}
\end{equation}
and
\begin{equation}
A(l)^{(d_l)}=\left(\begin{array}{cccc}0&0&0&\alpha\\
1&0&0&0\\
0&1&0&0\\
0&0&1&0
\end{array}
\right).\label{eq:ad}
\end{equation}
For $l<4$, the open boundary of the system at $l=1$ is incorporated by taking $A(l)^{(\sigma)}$ as the upper left
$(l+1)\times l$ block of the above defined matrices. We now apply this recursively, starting from
\begin{equation}
\left|\alpha\right>=A(L)^{(\sigma_L)}_{1j'}\left|j',\sigma_L\right>.
\end{equation}
Finally, we obtain the MPS expansion as
\begin{align}
&\left|\alpha\right>=\nonumber\\
&\left[A(L)^{(\sigma_L)}A(L-1)^{(\sigma_{L-1})}\ldots A(1)^{(\sigma_1)}\right]_{1,1}\left|\sigma_1\sigma_2\ldots\sigma_L\right>.\label{eq:mps}
\end{align}
Thus, $\left|\alpha\right>$ has bond dimension $4$. Even though it applies to the ground state of
the $t$-$V$ model at large $V$, it is hard to imagine being able to derive it from a perturbative
expansion. 

We now turn to another insightful question that cannot easily be answered from perturbative approaches,
namely how to find effective parent Hamiltonians that have asymptotically the same ground state 
as the $t$-$V$ model at large $V$. We provide three complementary answers to this problem in
the next section.
 
\section{Exact parent Hamiltonians}
\label{sec:parent}
We have seen that $\left|\alpha\right>$ is a Coupled Cluster state, that it admits a simple MPS
representation, and further that the calculation of expectation values can be performed by solving
combinatorial problems of rods in classical statistical physics. It turns out that each of these three
properties of $\left|\alpha\right>$ allows us to construct a distinct parent Hamiltonian for which it
is the exact ground state. We review each construction in turn, which highlights novel 
connections between seemingly distinct topics in many-body physics.

\subsection{Coupled Cluster parent Hamiltonian}
A first parent Hamiltonian containing only local terms, such that
$\left|\alpha\right>=\exp(\alpha T)\left|\text{MF}\right>$ is its unique ground state, can be constructed 
by exploiting the Coupled Cluster property of $\left|\alpha\right>$.
We start the construction by noting that the mean-field state $\left|\text{MF}\right>$ is the unique state 
that is annihilated by $c_{2j-1}$ and $c_{2j}^\dagger$, for all $j=1,\ldots,N$. This means that
$\left|\alpha\right>=\exp(\alpha T)\left|\text{MF}\right>$ is the unique state annihilated by
\begin{align}
&C_{2j-1}\equiv e^{\alpha T} c_{2j-1} e^{-\alpha T}=c_{2j+1}+\alpha Q_{2j-1},\nonumber\\
&C_{2j}\equiv e^{\alpha T} c_{2j}^\dagger e^{-\alpha T}=c_{2j}^\dagger+\alpha Q_{2j},\label{eq:simtrans1}
\end{align}
where 
\begin{eqnarray}
Q_{2j-1}&=&c_{2j-4}c_{2j-3}^\dagger c_{2j-2} - c_{2j-3}^\dagger c_{2j-2} c_{2j}\nonumber\\
&&+ c_{2j-2} c_{2j} c_{2j+1}^\dagger-c_{2j} c_{2j+1}^\dagger c_{2j+2}, \nonumber\\
Q_{2j}&=&c_{2j-3}^\dagger c_{2j-2} c_{2j-1}^\dagger-c_{2j-2}c_{2j-1}^\dagger c_{2j+1}^\dagger\nonumber\\
&&+c_{2j-1}^\dagger c_{2j+1}^\dagger c_{2j+2}-c_{2j+1}^\dagger c_{2j+2}c_{2j+3}^\dagger.\label{eq:simtrans2}
\end{eqnarray}
Note here that $C_{2j-1}$ and $C_{2j}$ are local operators in the Wannier basis associated with the
$c$ operators. Note further that $C_{2j}$ and $C_{2j-1}$ are obtained by a similarity transformation
that is not unitary. Hence, $\{C_n,C_{n'}^\dagger\}\not = \delta_{n,n'}$, i.e. the transformed
operators and their hermitian conjugates do not obey the canonical fermion creation-annihilation
operator algebra. Nonetheless, the fact that $\left|\alpha\right>$ is uniquely determined by
demanding that $C_n\left|\alpha\right>=0$, $n=1,\ldots,2N$, means that $\left|\alpha\right>$ is the
unique eigenstate of \begin{equation} H_\text{CC}\equiv\sum_{m=1}^{2N} C_m^\dagger
C_m,\label{eq:hpdef} \end{equation} with eigenvalue zero. Since $H_\text{CC}$ is semi-positive
definite, this means that $\left|\alpha\right>$ is the unique ground state of $H_\text{CC}$. 
The fact that $\left|\alpha\right>$ is simultaneously a zero-eigenstate of each local term $C_m^\dagger
C_m$ puts $H_\text{p}$ into a class of Hamiltonians that are called ``frustration free''. After
some algebra, $H_\text{p}$ is finally cast into the form:
\begin{align}
H_\text{CC}=&\sum_{j=1}^{N} \left(c_{2j-1}^\dagger c_{2j-1}+c_{2j}c_{2j}^\dagger\right) -4\alpha \left(T+T^\dagger\right)\nonumber\\
&+\alpha^2\sum_{j=1}^{N}\left(Q_{2j-1}^\dagger Q_{2j-1} + Q_{2j}^\dagger Q_{2j}\right). \label{eq:parent}
\end{align} 

Interestingly, the roles of single-particle and interaction terms are reversed compared to the
original $t$-$V$ Hamiltonian: the one-body term in~(\ref{eq:parent}) yields a flat dispersion
relation, and all hopping terms are
correlated. Flat dispersion and correlated hopping were previously encountered in an effective
description of fermions on a Creuz ladder~\cite{Mahyaeh2022}. Here these features emerge in another
(simpler) one-dimensional setting. At small $\alpha$, relevant for understanding correlations in
the insulating phase at $V>2t$ of the $t$-$V$ model, there is a hierarchy, with the $\alpha^0$ term,
which defines the mean-field configuration, being dominant. For sufficiently small $\alpha$, this
guarantees that $H_\text{CC}$ is gapped.

From $H_\text{CC}$, we see that three-body interactions are sufficient to stabilize the state
$\left|\alpha\right>$ for all $\alpha$. Here it is interesting to note a connection to
$\eta$-pairing~\cite{Yang1989} generalized to one-dimensional spinless fermions~\cite{Lorenzo2022}.
The concept refers to towers of exact scar eigenstates generated by a pairing operator schematically of the form $\sum
c_{j_1}^\dagger c_{j_2}^\dagger$ . They were first noticed in the Hubbard model. In Ref.~\cite{Lorenzo2022}, 
such states were stabilized in a one-dimensional spinless fermion model by three-body interactions. 
Larger $\sum c_{j_1}^\dagger c_{j_2}^\dagger \ldots c_{j_n}^\dagger$
multimers were also stabilized using $n$ body interactions. Our construction can be viewed as a
particle-hole transformed variant of $n=4$ tetramers that conserve particle number and that are 
stabilized as the ground state by $3$-body instead of $4$-body interactions.

\subsection{MPS parent Hamiltonian}
It is known than any MPS can be associated with a parent Hamiltonian involving local interactions
\cite{Fannes1992,Fernandez-Gonzalez2015}. However the states of the local Hilbert space do not necessarily
correspond to sites of the underlying lattice, but may be blocks of sites. The starting point is to consider $2^n$
strings $A^{\sigma_n}\ldots A^{\sigma_1}$ of the $4\times4$ matrices (\ref{eq:ar}) and (\ref{eq:ad})
that comprise the MPS representation. In the general theory~\cite{Fernandez-Gonzalez2015} the length
$n$ must be chosen such that the $2^n$ matrices that are generated by ranging over all possible
choices of $\sigma_1,\ldots,\sigma_n$ span the 16-dimensional space of $4\times 4$ matrices. This
means that $n\geq 4$. Brute force evaluations show that for the two matrices (\ref{eq:ar}) and
(\ref{eq:ad}), $n=6$. The resulting parent Hamiltonian then contains terms describing interactions
between adjacent six-site blocks. Since these terms have twelve-site footprints, the general
MPS construction leads to a cumbersome parent Hamiltonian, that we will not report explicitly
here.

\subsection{Rokshar-Kivelson parent Hamiltonian}
We finally devise a mapping to generalized quantum dimer models, usually of relevance for 
insulating RVB spin liquids~\cite{Moessner2001}. The precise connection can be built from quantum 
Hamiltonians that admit a Stochastic Matrix Decomposition~\cite{Castelnovo2005} 
and are dual to stochastic models in classical statistical physics. A preferred basis then exists for 
the quantum model, in which the expansion coefficients of the ground state are obtained by evaluating the
partition function of the classical model. Even in cases where the ground state has a simple MPS
structure, it may be very challenging to gain insight into excited states~\cite{Moudgalya2018}. It
is therefore remarkable that for systems with Stochastic Matrix Decomposition, equilibrium dynamic
correlators of the quantum model are dual the non-equilibrium stochastic Markovian dynamics of the
classical model. Typically, the Stochastic Matrix Decomposition requires a fine-tuning between the
diagonal and off-diagonal matrix elements in the preferred basis. The fine-tuned Hamiltonian is then
said to be at a Rokhsar-Kivelson point. The fact that calculations we performed for the state
$\left|\alpha\right>$ were equivalent to classical statistical physics calculations for a lattice
gas of four-site rods (tetramers), suggests that $\left|\alpha\right>$ may be associated to a 
Rokhsar-Kivelson parent Hamiltonian. Here we confirm this intuition.

We have to work in the Hilbert space accounting for all possible ways that the mean-field order can be
decorated with four-site disruptions. Formally we define the index set $\mathcal I$ as the set of
all sets of indices $\{j_1,\ldots,j_m\}$ with $0\leq 2m<N$, such that $1\leq j_l\leq 2N$ for each
index $j_l$, and furthermore $j_l\leq j_{l+1}-4$. For $J$ in $\mathcal I$, we define the state
\begin{equation} \left|J\right>=\prod_{j\in J} D_j\left|\text{MF}\right>.\label{eq:dstates}
\end{equation} Periodic boundary conditions are handled by letting $D_j$ in (\ref{eq:defd}) wrap
around past site $1$ to sites $2N-2,\ldots,2N$ for $j=1,2,3$. The set of states $\mathcal
B=\left\{\left|J\right>|J\in\mathcal I\right\}$ forms a basis for the space $\mathcal D$ of all
possible decorations of the mean-field order by four-site disruptions.

Next we define local diagonal operators that will eventually be associated with potential energy
terms in the parent Hamiltonian. The first projects out the component of a state in the $\mathcal
B$ basis for which the five sites $j-4,\ldots,j$ are of regular order
$\cdots\circ\circ\circ\circ\circ\cdots$:
\begin{equation}
\left|\circ\circ\circ\circ\circ\right>\left<\circ\circ\circ\circ\circ\right|_j=\sum_{\{\sigma\}_j}\left|\{\sigma\}_j\right> \left<\{\sigma\}_j\right|,\label{eq:r5}
\end{equation}
where
\begin{equation}
\left|\{\sigma\}_j\right>=\left|\sigma_1\ldots\sigma_{j-5}r_{j-4}\ldots r_j\sigma_{j+1}\ldots\sigma_{2N}\right>,
\end{equation}
(We find it convenient here to denote sites of regular order with open circles and disrupted
sites with dots. Note that this is different from the convention used in Figure~\ref{fig:state},
where
open/filled circles indicated the occupation of orbitals.) 
The second and third projects out components in the $\mathcal B$ basis in which a string of
head-to-toe four-site disruption starts on site $j-3$ or ends on site $j-1$. Here and below
``start'' or ``end'' refers to where the string is bordered on respectively the left or right by a
site of regular order 
$\cdots \circ\bullet\bullet\bullet\bullet\cdots$ or $\cdots \bullet\bullet\bullet\bullet\circ\cdots$.
\begin{eqnarray}
\left|\circ\bullet\bullet\bullet\bullet\right>\left<\circ\bullet\bullet\bullet\bullet\right|_j&=&D_j\left|\circ\circ\circ\circ\circ\right>\left<\circ\circ\circ\circ\circ\right|_jD_j^\dagger,\nonumber\\
\left|\bullet\bullet\bullet\bullet\circ\right>\left<\bullet\bullet\bullet\bullet\circ\right|_j&=&D_{j-1}\left|\circ\circ\circ\circ\circ\right>\left<\circ\circ\circ\circ\circ\right|_jD_{j-1}^\dagger.\nonumber\\
\end{eqnarray}
Finally, with the aid of (\ref{eq:r5}) and $D_j$, we define four off-diagonal operators that will be associated with kinetic terms in the
parent Hamiltonian. They create or destroy a four-site disruption starting at $j-3$ when $j-4$ is of
regular order $\cdots\circ\circ\circ\circ\circ\cdots\leftrightarrow \cdots
\circ\bullet\bullet\bullet\bullet \cdots$, or ending at $j-1$ if $j$ is of regular order
$\cdots\circ\circ\circ\circ\circ\cdots\leftrightarrow \cdots\bullet\bullet\bullet\bullet\circ\cdots$. 
\begin{eqnarray}
\label{eq:odops}
\left|\circ\bullet\bullet\bullet\bullet\right>\left<\circ\circ\circ\circ\circ\right|_j&=&D_{j}\left|\circ\circ\circ\circ\circ\right>\left<\circ\circ\circ\circ\circ\right|,\\
\left|\circ\circ\circ\circ\circ\right>\left<\circ\bullet\bullet\bullet\bullet\right|_j&=&\left|\circ\circ\circ\circ\circ\right>\left<\circ\circ\circ\circ\circ\right|_jD_j^\dagger,\nonumber\\
\left|\bullet\bullet\bullet\bullet\circ\right>\left<\circ\circ\circ\circ\circ\right|_j&=&D_{j-1}\left|\circ\circ\circ\circ\circ\right>\left<\circ\circ\circ\circ\circ\right|,\nonumber\\
\left|\circ\circ\circ\circ\circ\right>\left<\bullet\bullet\bullet\bullet\circ\right|_j&=&\left|\circ\circ\circ\circ\circ\right>\left<\circ\circ\circ\circ\circ\right|_jD_{j-1}^\dagger.\nonumber
\end{eqnarray}
Note that the operators (\ref{eq:odops}) can be used to connect any state in $\mathcal D$ to any
other state in $\mathcal D$: they allow one to add or remove a four-site disruption from either end
of a string or head-to-toe disruptions. Combining two such operators allows us to move any four-site
disruption whenever it is bordered by a site of regular order. We call this property
``connectedness''.

In the space $\mathcal D$ of all possible decorations of the mean-field order by four-site disruptions, we define the Rokhsar Kivelson parent Hamiltonian
\begin{align}
H_\text{RK}=& \sum_{j=1}^{2N}\Big[ 2\alpha^2\left|\circ\circ\circ\circ\circ\right>\left<\circ\circ\circ\circ\circ\right|_j\nonumber\\
&+\left|\circ\bullet\bullet\bullet\bullet\right>\left<\circ\bullet\bullet\bullet\bullet\right|_j
+\left|\bullet\bullet\bullet\bullet\circ\right>\left<\bullet\bullet\bullet\bullet\circ\right|_j\nonumber\\
&-\alpha \Big(\left|\circ\bullet\bullet\bullet\bullet\right>\left<\circ\circ\circ\circ\circ\right|_j+\left|\circ\circ\circ\circ\circ\right>\left<\circ\bullet\bullet\bullet\bullet\right|_j\nonumber\\
&+\left|\bullet\bullet\bullet\bullet\circ\right>\left<\circ\circ\circ\circ\circ\right|_j+\left|\circ\circ\circ\circ\circ\right>\left<\bullet\bullet\bullet\bullet\circ\right|_j\Big)\Big]\nonumber\\
=&\sum_{j=1}^{2N} (P_{1,j}+P_{2,j}),\label{eq:hrk}
\end{align}
where
\begin{align}
P_{1,j}&=\big(\alpha-D_j\big)\left|\circ\circ\circ\circ\circ\right>\left<\circ\circ\circ\circ\circ\right|_j\big(\alpha-D_j^\dagger\big)\nonumber\\
P_{2,j}&=\big(\alpha-D_{j-1}\big)\left|\circ\circ\circ\circ\circ\right>\left<\circ\circ\circ\circ\circ\right|_j\big(\alpha-D_{j-1}^\dagger\big).
\end{align}
This provides a phenomenological model for fluctuations of mean-field order generated by tetramers. There is a potential cost $1$ associated with the ends
$\circ\bullet\bullet\bullet\bullet$ and $\bullet\bullet\bullet\bullet\circ$ of strings of
head-to-toe disruptions, but no energy cost associated with the bulk of the string. The kinetic
terms of order $\alpha$ create new disruptions or grow/shrink the ends of existing ones 
(by four sites at a time). The fine-tuning characteristic of Rokhsar-Kivelson points amounts to a potential
energy cost $2\alpha^2$ associated with not having disruptions in a region
$\circ\circ\circ\circ\circ$ that could host them.

The connectedness property of the operators used to construct it, implies that $H_\text{RK}$ is irreducible with respect to the basis $\mathcal B$:
Every pair of elements in $\mathcal B$ have a non-zero matrix element for some power of $H_\text{RK}$. 
Note further that for an appropriate choice of $\Lambda$, 
all the matrix elements of $I\Lambda-H_\text{RK}$ are non-negative. The Perron-Frobenius theorem~\cite{Meyer2000} then guarantees that $H_\text{RK}$ has a non-degenerate ground state.
$H_{RK}$ is the sum of semi-positive-definite terms. As a result, its ground state energy is bounded from below by zero.
It is straightforward to show that the non-degenerate ground state is $\left|\alpha\right>=e^{\alpha T}\left|\text{MF}\right>$ and that it has zero energy.

We start by writing $\left|\alpha\right>=(1+\alpha D_j)\left|\alpha,j\right>$. Note that $\left|\alpha,j\right>=e^{\alpha\sum_{j'\not=j}D_{j'}}\left|\text{MF}\right>$ is formed without ever acting with $D_j$. As a result, none of the basis elements from $\mathcal B$ that make up $\left|\alpha,j\right>$
contains a string of head-to-toe disruptions ending on site $j$ or starting on site $j-3$.
Then
\begin{align}
&\left|\circ\circ\circ\circ\circ\right>\left<\circ\circ\circ\circ\circ\right|_j\big(\alpha-D_j^\dagger\big)\left|\alpha\right>\nonumber\\
=&\alpha\left|\circ\circ\circ\circ\circ\right>\left<\circ\circ\circ\circ\circ\right|_j\left|\alpha,j\right>\nonumber\\
&-\alpha \underbrace{\left|\circ\circ\circ\circ\circ\right>\left<\circ\circ\circ\circ\circ\right|_jD_j^\dagger D_j}_1 \left|\alpha,j\right>\nonumber\\
&+\alpha^2 \underbrace{\left|\circ\circ\circ\circ\circ\right>\left<\circ\circ\circ\circ\circ\right|_j D_j}_2 \left|\alpha,j\right>\nonumber\\
&- \underbrace{\left|\circ\circ\circ\circ\circ\right>\left<\circ\bullet\bullet\bullet\bullet\right|_j \left|\alpha,j\right>}_3.
\end{align}
When acting to the left $D_j^\dagger D_j$ adds and then removes the same disruption from the regular sites $j-3,\ldots,j$ in the part marked (1), so that
$\left|\circ\circ\circ\circ\circ\right>\left<\circ\circ\circ\circ\circ\right|_jD_j^\dagger D_j=\left|\circ\circ\circ\circ\circ\right>\left<\circ\circ\circ\circ\circ\right|_j$. The part marked (2) is zero because when acting to the left,
$D_j$ tries to remove a disruption from states that don't have any disruptions on sites $j-3,\ldots,j$. 
The part marked (3) is zero because $\left|\circ\circ\circ\circ\circ\right>\left<\circ\bullet\bullet\bullet\bullet\right|_j$ looks for head-to-toe strings of disruptions that start on site $j-3$ and $\left|\alpha,j\right>$ has no such terms.
Thus $P_{1,j}\left|\alpha\right>=0$. When $\left|\alpha\right>$ is written $\left|\alpha\right>=(1+\alpha D_j)\left|\alpha,j\right>$, a similar argument applies for $P_{2,j+1}\left|\alpha\right>$.
The choice of $j$ was arbitrary and hence $H_{RK}\left|\alpha\right>=0$.

It is immediately apparent from (\ref{eq:hrk}) that $H_{RK}$ is of the Stochastic Matrix Form. See
Eq.~(15a) and (15b) in~\cite{Castelnovo2005}. 
The classical model that $H_{RK}$ is dual to is a lattice gas of tetramers (four-site rods) in one
dimension, with chemical potential $\beta\mu=2\ln\alpha$, but no kinetic energy. This is the physics
relevant for linear adsorbates~\cite{Ramirez-Pastor1999}. Here the system is endowed with Markovian
stochastic dynamics in the Grand Canonical Ensemble involving random transitions 
$\cdots\circ\circ\circ\circ\circ\cdots\leftrightarrow \cdots \circ\bullet\bullet\bullet\bullet \cdots$ and
$\cdots\circ\circ\circ\circ\circ\cdots\leftrightarrow \cdots\bullet\bullet\bullet\bullet\circ\cdots$. 

The parent Rokhsar-Kivelson Hamiltonian $H_\text{RK}$ clearly differs from the parent Coupled Cluster 
Hamiltonian $H_\text{CC}$ of Eq.~(\ref{eq:parent}).
Local terms in $H_\text{CC}$ have a footprint of up to 7 sites, as opposed 5 sites for
$H_\text{RK}$. Furthermore, there are terms at order $\alpha^2$ in $H_\text{CC}$ that move disruptions around, 
whereas in $H_\text{RK}$, moving a disruption is a second-order process first encountered in
$(H_\text{RK})^2$. The translation between the different languages (Coupled Cluster state, matrix
product state, Rokhsar-Kivelson state) describing the post-Hartree-Fock charge insulator is thus non-trivial.

\section{Non-perturbative Coupled Cluster state}
\label{sec:lalpha}
Up to this point, we have viewed the Coupled Cluster state $\left|\alpha\right>$ as an
asymptotic description of the ground state of the $t$-$V$ model for small $t/V$, which meant 
that we were only interested in the small $\alpha$ regime of the ansatz.
However, as we argued in the previous section, the parent Hamiltonians that we identified there are
of physical interest in their own right. When viewed as the ground state of such a parent Hamiltonian, 
the restriction that $\alpha$ should be small can be lifted. We set here the final goal to obtain 
the covariance matrix in the $c_j$ basis at arbitrary $\alpha$ and in the thermodynamic limit. 
From Eq.~(\ref{eq:twoc}), we have:
\begin{equation}
\left<\alpha\right|c_1^\dagger c_{1+4l}\left|\alpha\right>
=-\alpha^{2l}\sum_{j=0}^\infty {2 N - 4l -1 -3j \choose j} \alpha^{2j},\label{eq:cc1}
\end{equation}
where we define ${n\choose j}=0$ for $n<0$ and we take $l>0$. We can convert the sum into a contour integral:
\begin{align}
&\left<\alpha\right|c_1^\dagger c_{1+4l}\left|\alpha\right>=-\alpha^{2l}\sum_{n=0}^\infty \sum_{j=0}^n {n \choose j} \delta_{2N-4l-1-3j,n}\alpha^{2j}\nonumber\\
&=-\alpha^{2l}\int_0^{2\pi}\frac{d\phi}{2\pi}e^{-(2N-4l-1)\phi}\\
&~~~~~~~~~~~~~~~~~\times\sum_{n=0}^\infty\sum_{j=0}^n {n \choose j}\left(e^{4i\phi}\alpha^2\right)^j\left(e^{i\phi}\right)^{n-j}\nonumber\\
&=-\alpha^{2l}\int_0^{2\pi}\frac{d\phi}{2\pi}e^{-(2N-4l-1)\phi}\sum_{n=0}^\infty \left(e^{i\phi}+\alpha^2 e^{4i\phi}\right)^n\nonumber\\
&=-\alpha^{2l}\int_0^{2\pi}\frac{d\phi}{2\pi}\frac{e^{-(2N-4l-1)\phi}}{1-e^{i\phi}-\alpha^2e^{4i\phi}}\nonumber\\
&=-\alpha^{2l}\oint_0^{2\pi}\frac{d\zeta}{2\pi i}\frac{1}{\zeta^{2N-4l}(1-\zeta-\alpha^2\zeta^4)},
\end{align}
where the integral traverses the unit circle in the complex plane in an anti-clockwise sense. When
applying the residue theorem to this contour integral, we encounter two types of poles. There is the
high-order pole at $\zeta=0$, which leads to a residue
\begin{equation}
\left.\frac{1}{(2N-4l-1)!}\partial_\zeta^{2N-4l-1}\frac{1}{1-\zeta-\alpha^2 \zeta^4}\right|_{\zeta=0}.
\end{equation}
It gives an $\mathcal O(N^0)$ contribution in the thermodynamic limit, owing to the factorial in the
denominator. This contribution is overwhelmed by the residue at roots of $1-\zeta-\alpha^2 \zeta^4$
inside the unit circle. In the thermodynamic limit and to exponential accuracy in $N$ then, 
\begin{align}
\left<\alpha\right|c_1^\dagger c_{1+4l}\left|\alpha\right>=-\frac{(\alpha^2 \zeta_0^4)^l}{\zeta_0^{2N}}\lim_{\zeta\to \zeta_0}\frac{\zeta-\zeta_0}{1-\zeta-\alpha^2 \zeta^4},
\end{align}
where $\zeta_0$ is the real root of $1-\zeta-\alpha^2\zeta^4$ inside the unit circle that is closest to the origin.
An explicit expression is provided in Appendix~\ref{app:zeta}. The root $\zeta_0(\alpha)$
is a monotonically decreasing function of $\alpha$. In the small $\alpha$ regime,
$\zeta_0=1-\alpha^2+\ldots$ whereas at large $\alpha$, $\zeta_0\simeq1/\sqrt{\alpha}$.

We can use the same method as above to obtain
\begin{equation}
\left<\alpha\right|\left.\alpha\right>=\frac{1+3\alpha^2\zeta_0^4}{\zeta_0^{2N+1}}\lim_{\zeta\to \zeta_0}\frac{\zeta-\zeta_0}{1-\zeta-\alpha^2 \zeta^4}.
\end{equation}
This then gives in the thermodynamic limit
\begin{equation}
\frac{\big<\alpha|c_1^\dagger c_{1+4l}|\alpha\big>}{\left<\alpha\right|\left.\alpha\right>}
=-\frac{\alpha^{2l}\zeta_0^{4l+1}}{1+3\alpha^2\zeta_0^4}=-\frac{(1-\zeta_0)^l \zeta_0}{4(1-\zeta_0)+\zeta_0}.\label{eq:covther}
\end{equation}
From (\ref{eq:diag}) follows that $\big<\alpha|c_2^\dagger c_2|\alpha\big>/\left<\alpha\right|\left.\alpha\right>$ can be obtained from
(\ref{eq:covther}) by multiplying by $-1$ and setting $l=0$. Since 
$\big<\alpha|c_1^\dagger c_1|\alpha\big>/\left<\alpha\right|\left.\alpha\right>=1-\big<\alpha|c_2^\dagger c_2|\alpha\big>/\left<\alpha\right|\left.\alpha\right>$, 
we then find for all $l$ in the thermodynamic limit
\begin{equation}
\frac{\big<\alpha|c_1^\dagger c_{1+4l}|\alpha\big>}{\left<\alpha\right|\left.\alpha\right>}=\delta_{l,0}-\frac{(1-\zeta_0)^{|l|} \zeta_0}{4(1-\zeta_0)+\zeta_0}.\label{eq:corfin}
\end{equation}
The full covariance matrix $Q_{jj'}=\big<c_j^\dagger c_{j'}\big>$ is deduced from the 2-site translation
symmetry of the charge ordered state.
It is interesting to compare this exponential decay for the Coupled Cluster state to the decay
found for the critical system at $V=-2t$. In both cases, an ordered state ($\left|\text{MF}\right>$
in the former and the fermionic vaccum $\left|0\right>$ in the latter case) is decorated with a
finite density of randomly placed disruptions. For the Coupled Cluster state, the disruptions $D_j$ contain four
adjacent fermion operators, while for the critical state, the disruptions are single fermions. 
Randomness on its own causes an exponential decay of single-particle correlations with a finite decay length
(\ref{eq:corfin}). The additional destructive interference caused by the fermionic sign fluctuations
when the disruptions are single fermions, cause exponential decay with a vanishing decay length 
$1/\ln N$ at large $N$, see (\ref{eq:Qexact}).

\begin{center}
\begin{figure}
\includegraphics[width=0.97\columnwidth]{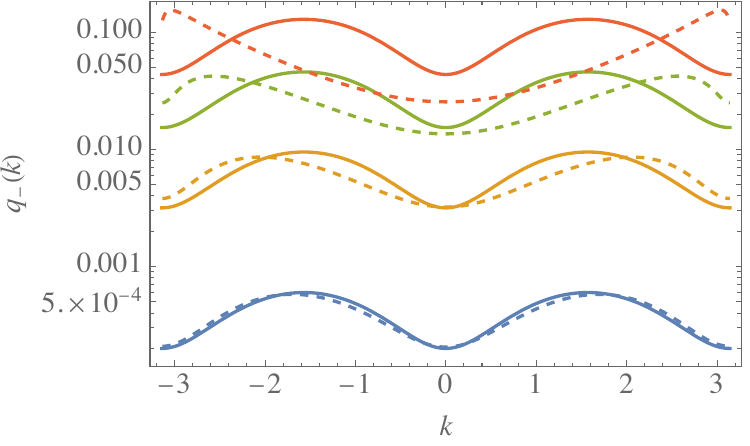}\\
\hspace{.075\columnwidth}\includegraphics[width=0.92\columnwidth]{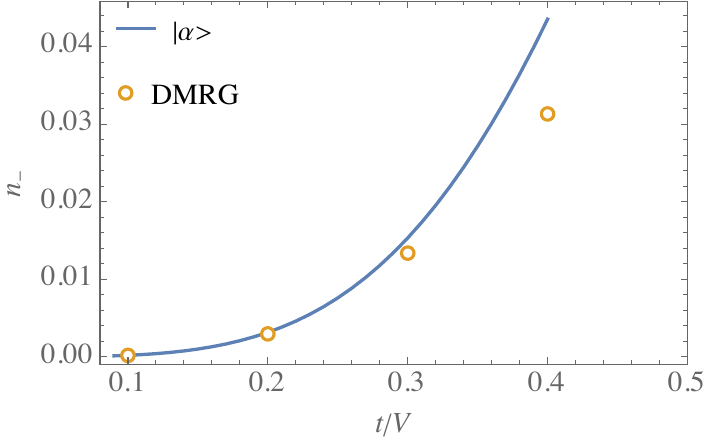}
\caption{\label{fig:lalphaspec} Top panel: comparison of the numerically exact DMRG results (dashed
curves) for the lower band of the covariance matrix spectrum of the $t$-$V$ model, to (the Fourier transform of) 
the result (\ref{eq:corfin}) obtained for the non-perturbative Coupled Cluster state (solid curves). 
The value of $t/V=\sqrt{\alpha}$ for curves from bottom to top are $t/V=0.1,\,0.2,\,0.3,\,0.4$. 
Bottom panel: The density $n_-$ of quasiparticles associated with the upper band of the mean-field Hamiltonian versus $t/V$.} 
\end{figure}
\end{center}

While we have taken the view in this section that $\left|\alpha\right>$ with $\alpha>0.01$ is an
interesting state in its own right, we conclude this section by pointing out that this state,
although not quantitatively accurate, still 
captures reasonably well the extent of correlations in the $t$-$V$ model at $t/V>0.1$. In the top
panel of Figure~\ref{fig:lalphaspec} we compare the lower band of the covariance matrix spectrum at
values of $t/V$ between $0.1$ and $0.4$, obtained from numerically exact DMRG calculations, to the
result obtained by Fourier transforming (\ref{eq:corfin}) with $\alpha=(t/V)^2$. We see that the
average position and bandwidth agree rather well between the two calculations, although
the progressive change in periodicity is of course not captured by the Coupled Cluster state. 

In the bottom panel of Fig.~\ref{fig:lalphaspec}, we also compare results for the average density of the quasiparticles 
associated with the upper band of the mean-field Hamiltonian, i.e.
\begin{equation}
n_-=\big<c_1^\dagger c_1\big>/2.
\end{equation}
(The factor $1/2$ reflects the fact that the unit cell contains two chain sites.) We see good
agreement that is not automatically guaranteed. For instance, the post-Hartree-Fock treatment
in~\cite{Gebhard2022} cannot reproduce this agreement. (See Figures 12 and 13 in that work.) 
While the actual disruption structure at $V\gtrsim 2t$ becomes more complicated than the simple 
four-site patten of $\left|\alpha=(t/V)^2\right>$, the latter state nonetheless still contains 
roughly the right amount of correlations, and constitute a natural physical picture of correlated
insulators.

\section{Conclusions and perspectives}
\label{sec:conclusions}
In this work, we have uncovered a simple Coupled Cluster state, given by Eq.~(\ref{eq:ansatz}), that
asymptotically describes the correlated charge insulating ground state of the $t$-$V$ model in one dimension 
at large $V/t$. This emergent structure remains hidden in usual exact treatments of one-dimensional models, 
such as the Bethe ansatz or the density matrix renormalization group.
Our result was obtained by bootstrapping from a perturbative calculation of the covariance matrix
spectrum (\ref{eq:qk}) to a Coupled Cluster state that variationally reproduces the ground state
energy to one order further than the Hartree-Fock mean-field theory of the correlated charge insulator.
This enabled us to answer analytically questions that cannot be addressed by perturbative means. We showed
that the leading order fluctuations beyond mean-field are due to random decorations of the ordered
state by four-site disruptions of opposite orientation to the underlying mean-field charge pattern 
(schematically illustrated in Figure~\ref{fig:state}). 
We also showed that the four-site disruptions state leads to a Matrix Product state with
bond-dimension exactly four (\ref{eq:mps}). While there is a general construction~\cite{Fannes1992}
that relates Matrix Product states to parent Hamiltonians with local terms only, that route does not
seem to yield a parent Hamiltonian of much pratical use in this case. We have instead derived two
distinct parent Hamiltonians. One parent Hamiltonian, given by Eq.~(\ref{eq:hrk}), constitutes a phenomenological 
model for quantum fluctuations induced by tetramers (length four rods), that connects the quantum system 
to the classical stochastic dynamics of linear adsorbates~\cite{Ramirez-Pastor1999}
 when tuned to its Rokhsar-Kivelson point~\cite{Castelnovo2005}.
The other parent Hamiltonian, given by Eq.~(\ref{eq:parent}), is a correlated hopping
Hamiltonian involving up to only three-body interactions. We have also obtained exact results for
the covariance matrix (\ref{eq:corfin}) of the Coupled Cluster state in the thermodynamic limit, and
showed that the level of correlation is similar to that of the $t$-$V$ model even in a regime where
there are quantitative differences between the Coupled Cluster and exact ground states.

Our results make the case for further studies of the covariance matrix of systems in which ground
states have non-trivial unit-cell structure. We briefly speculate here regarding potential future
developments. One interesting question is how to break free from the four-sites periodic
structure of the Coupled Cluster state based on four-sites disruptions. This is not only required 
to improve agreement at larger values of $t/V$ in the correlated insulator phase, but also to understand 
which are the new physical processes at play in this regime.
Another promising avenue is to use the Coupled Cluster state structure that we uncovered as
a starting point to build more intricate many-body states in one dimension. 
A direct and interesting extension of our work would be to consider one-dimensional models of
Mott insulators with spinful fermions, as well as ladders, and to investigate the emergent structures
in the presence of spin and charge degrees of freedom.
In addition, since the exact gapless critical state was found to show some similarities to the
gapped Coupled Cluster state, it is conceivable that further extension of our work could lead to 
an explicit construction of correlated wave functions for the Luttinger liquid phase.
Studies of correlated phases of charge insulators beyond one dimension~\cite{Menczer2024} could also be considered, 
using the mapping to generalized quantum multi-mer models.
Another avenue to be explored relates to topologically non-trivial phases of interacting systems. For correlated 
systems with non-trivial unit-cell structure, the covariance matrix in $k$-space may play a similar role to a
single-particle Hamiltonian, thus allowing one to borrow from the theory of topological band
insulators \cite{Hannukainen2024}. In this context, there are possibly subtle but interesting questions about if and how the
bulk-edge correspondence of the band insulator theory carries over to the theory of the covariance
matrix of a correlated insulator. 

\acknowledgements{This work was supported by European Research Council (ERC) under Grant Agreement No. 101001310 466 (SuperProtected). IS thanks the CNRS for financial mobility support.}

\appendix
\section{Exact ground state at $\mathbf{V=-2t}$}
\label{app:GS_crit}
In the appendix of~\cite{Barghathi2019}, it was proven that (\ref{eq:GS_crit}) is the ground state of
the $t$-$V$ model at $V=-2t$. The proof was performed in the fermionic language. Here we sketch the
equivalent proof in the spin-chain language, which is related to the fermion language by the
Wigner-Jordan mapping to spin-1/2 operators 
\begin{equation}
S_{j+}=S_{xj}+iS_{yj}=e^{i\pi\sum_{l=1}^{j-1} a_l^\dagger a_l} a_j^\dagger\label{eq:splus}
\end{equation}
that commute if $j\not=j'$.
In this representation the Hamiltonian at $V=-2t$ corresponds to the spin-isotropic, ferromagnetic 
Heisenberg spin chain:
\begin{equation}
H=-\frac{t}{2}\sum_{j=1}^{2N}\bm S_j\cdot \bm S_{j+1}.
\end{equation}
The total spin and its $z$-component are good quantum numbers. The fully polarized ground state 
is the state $\left|J=N,J_z=-N\right>$, which corresponds to the empty state $\left|0\right>$ in the fermionic
language. The Hamiltonian commutes with $S_+=\sum_{j=1}^{2N} S_{j+}$, and thus the ground state 
with spin projection $J_z=0$ is given exactly by:
\begin{equation}
\left|J=N,J_z=0\right>\propto (S_+)^N\left|J=N,J_z=-N\right>.
\end{equation}
In the fermion language, this leads to the half-filled state (\ref{eq:GS_crit}), as is easy to check using 
the mapping (\ref{eq:splus}).

\section{Covariance matrix at $\mathbf{V=-2t}$}
\label{app:crit}
Here we perform the sum over combinatorial factors 
\begin{equation}
S=\sum_{n=0}^{2r+1}(-1)^n{2r \choose n}{2N-2r-2 \choose N-n-1},
\end{equation}
that produces the result (\ref{eq:Qexact}).
It is straightforward to see that 
\begin{align}
&{2r \choose n}{2N-2r-2 \choose N-n-1}\nonumber\\
&=\frac{(2r)!(2N-2r-2)!}{[(N-1)!]^2}{N-1\choose n}{N-1 \choose 2r-n},
\end{align}
so that
\begin{align}
&S=\frac{(2r)!(2N-2r-2)!}{[(N-1)!]^2}\underbrace{\sum_{n=0}^{2r+1}(-1)^n{N-1 \choose n}{N-1 \choose 2r-n}}.
\end{align}
The part marked with an underbrace is easily identified as the coefficient of the monomial $x^{2r}$ 
in the polynomial $(1-x)^{N-1}(1+x)^{N-1}=(1-x^2)^{N-1}$.
Thus by the binomial theorem
\begin{eqnarray}
S&=&\frac{(2r)!(2N-2r-2)!}{[(N-1)!]^2}{N-1 \choose r}\nonumber\\
&=&(-1)^r\frac{(2r)!(2N-2r-2)!}{r!(N-r-1)!(N-1)!}\nonumber\\
&=&(-1)^r\frac{2^{2N-2}\Gamma(r+1/2)\Gamma(N-r-1/2)}{\Gamma(1/2)^2(N-1)!}
\end{eqnarray}
which decays like $N^{-r}$ at large $N$, due to the $r$ unbalanced factors of order $N$ in the
denominator of $(N-r-3/2)(N-r-5/2)\ldots(1/2) /(N-1)(N-2)\ldots(1)$.

\section{Covariance matrix at order $\mathbf{(t/V)^4}$}
\label{app:covmat}
In the main text we present the covariance matrix spectrum of the $t$-$V$ model at small $t/V$, to
order $(t/V)^4$. It is obtained from the following expansion of the covariance matrix, to order
$(t/V)^4$ in $k$-space.
\begin{align}
&Q(k)=\frac{1}{2}\sigma_0\nonumber\\
&+\left[\left(1+\cos k\right)\frac{t}{V}-\left(2+5\cos k+3\cos 2k\right)\left(\frac{t}{V}\right)^3\right]\sigma_x\nonumber\\
&+\left[\sin k \frac{t}{V}+\left(\sin k -3 \sin 2k\right)\left(\frac{t}{V}\right)^3\right]\sigma_y\nonumber\\
&+\Bigg[-\frac{1}{2}+2(1+\cos k)\left(\frac{t}{V}\right)^2\nonumber\\
&~~~~~~~~~~~+\left(2-12\cos k - 12 \cos 2k\right)\left(\frac{t}{V}\right)^4\Bigg]\sigma_z.\label{eq:Qk}
\end{align} 
As mentioned in the main text, the spectrum of $Q(k)$ has periodicity $\pi$ whereas $Q(k)$ itself only has periodicity $2\pi$.

\section{Mean-field solution of the $\mathbf{t}$-$\mathbf{V}$ model}
\label{app:mf}
We seek a single particle basis
\begin{equation}
b_\alpha^\dagger=\sum_{j=1}^{2N} f_{j\alpha}a_j^\dagger,
\end{equation}
such that for a particular sub-set $\mathcal F$ of $N$ of these orbitals, the state
\begin{equation}
\left(\prod_{\alpha\in\mathcal F}b_\alpha^\dagger\right)\left|0\right>,
\end{equation}
minimizes the expectation value of the Hamiltonian (\ref{eq:tvham}), within the manifold of single
Slater determinants. Standard Hartree-Fock theory leads to a mean-field Hamiltonian
\begin{equation}
\varepsilon_\alpha f_{j\alpha}=\sum_{k=1}^{2N}h_{jk}f_{k\alpha},
\end{equation}
for the mean-field orbitals, where
\begin{align}
h_{jk}=&-\left(t+\big<a_{j+1}^\dagger a_j\big>\right)\delta_{k,j+1}
-\left(t+\big<a_{k}^\dagger a_{k+1}\big>\right)\delta_{j,k+1}\nonumber\\
&+V\left(\big<a^\dagger_{j-1}a_{j-1}\big>+\big<a^\dagger_{j+1}a_{j+1}\big>\right)\delta_{j,k}.\label{eq:mfham}
\end{align}
Periodic boundary conditions here imply $2N+1\equiv 1$ and $0\equiv 2N$. We seek an
inversion-symmetric solution that is invariant under translation by two lattice sites. We therefore
take $\big<a_j^\dagger a_j\big>=\big<a_{j+2}^\dagger a_{j+2}\big>$, and 
$\big<a_{j+1}^\dagger a_j\big>=\big<a_{j-1}^\dagger a_{j}\big>$ for all $j$. We further assume 
$\big<a_{j+1}^\dagger a_j\big>$ to be real. 

This results in two bands with $N$ orbitals each. For the lower band
\begin{equation}
f_{2j-1,\alpha,-}=\frac{x_\alpha}{\sqrt{N}} e^{ik_\alpha j},
~f_{2j,\alpha,-}=\frac{y_\alpha}{\sqrt{N}} e^{ik_\alpha j},
\end{equation}
with $k_\alpha=2\pi \alpha/N$, and $\alpha=-\frac{N-1}{2},\ldots,\frac{N-1}{2}$, $N$ being assumed odd.
The coefficients $x_{\alpha}$ and $y_{\alpha}$ form the negative-energy eigenstate of the eigenvalue equation
\begin{equation}
-\varepsilon_\alpha \left(\begin{array}{c} x_\alpha \\ y_\alpha \end{array} \right) = 
\left(\begin{array}{cc} \Delta & -\tilde t\left(1+e^{-ik_\alpha}\right)\\ 
-\tilde t\left(1+e^{ik_\alpha}\right) & -\Delta\end{array}\right) \left(\begin{array}{c} x_\alpha \\ y_\alpha \end{array} \right),
\end{equation}
where the mean-field parameters
\begin{eqnarray}
\Delta&=&V\left(\left<a_{2j}^\dagger a_{2j}\right>-\left<a_{2j-1}^\dagger a_{2j-1}\right>\right),\nonumber\\
\tilde t&=&t+V\left<a_{j+1}^\dagger a_j\right>,\label{eq:deltaandtt}
\end{eqnarray}
do not depend on $j$. 
From this we find:
\begin{eqnarray}
\varepsilon_\alpha&=&\sqrt{\Delta^2+4 \tilde t^2\cos^2\frac{k_\alpha}{2}},\\
x_\alpha&=&\frac{1}{\sqrt{2}}\sqrt{1-\frac{z}{\sqrt{z^2+\cos^2\frac{k_\alpha}{2}}}}e^{-ik_\alpha/2},\label{eq:defx}\\
y_\alpha&=&\frac{1}{\sqrt{2}}\sqrt{1+\frac{z}{\sqrt{z^2+\cos^2\frac{k_\alpha}{2}}}}.\label{eq:defy}
\end{eqnarray}
where $z=\Delta/2\tilde t$.
The upper band is described by orbitals:
\begin{equation}
f_{2j-1,\alpha,+}=\frac{y_\alpha}{\sqrt{N}} e^{ik_\alpha j},
~f_{2j,\alpha,+}=-\frac{x_\alpha^*}{\sqrt{N}} e^{ik_\alpha j}.
\end{equation}

To obtain self-consistency, we evaluate
\begin{eqnarray}
\big<a_{2j}^\dagger a_{2j}\big>-\big<a_{2j-1}^\dagger a_{2j-1}\big>&=&
\frac{1}{N}\sum_{\alpha=-\frac{N-1}{2}}^{\frac{N-1}{2}}\left(y_\alpha^2-|x_\alpha|^2|\right)\nonumber\\
&=&\frac{1}{N}\sum_{\alpha=-\frac{N-1}{2}}^{\frac{N-1}{2}}\frac{\Delta}{\varepsilon_\alpha},\label{eq:diagas}\\
\big<a_{2j}^\dagger a_{2j-1}\big>&=&\frac{1}{N}\sum_{\alpha=-\frac{N-1}{2}}^{\frac{N-1}{2}}x_\alpha y_\alpha\nonumber\\
&=&\frac{1}{N}\sum_{\alpha=-\frac{N-1}{2}}^{\frac{N-1}{2}}\frac {\tilde t \cos^2\frac{k_\alpha}{2}}{\varepsilon_\alpha},
\end{eqnarray}
that are independent of $j$. We also see that
\begin{eqnarray}
\big<a_{2j+1}^\dagger a_{2j}\big>&=&\frac{1}{N}\sum_{\alpha=-\frac{N-1}{2}}^{\frac{N-1}{2}}x_\alpha^* e^{-ik_\alpha} y_\alpha \nonumber\\
&=&\frac{1}{N}\sum_{\alpha=-\frac{N-1}{2}}^{\frac{N-1}{2}}x_\alpha y_\alpha=\big<a_{2j}^\dagger a_{2j-1}\big>
\end{eqnarray}
which confirms that it is consistent to take $\big<a_{j+1}^\dagger a_{j}\big>$ independent of $j$ and real.
All the above leads to a transcendental equation
\begin{equation}
\frac{t}{V}=\frac{1}{2N}\sum_{\alpha=-\frac{N-1}{2}}^{\frac{N-1}{2}}\frac{\sin^2\frac{k_\alpha}{2}}{\sqrt{z^2+\cos^2\frac{k_\alpha}{2}}},\label{eq:findz}
\end{equation}
which determines $z$. Without loss of generality, we pick the $z>0$ solution, which corresponds to a symmetry broken state in
which the even sites are more likely to be occupied than the odd ones. 

The mean-field orbitals that we have found are Bloch states of the reduced symmetry mean-field
Hamiltonian (\ref{eq:mfham}). Since we are studying an insulating state, a description in terms of
localized orbitals is however more useful. To obtain such a description, we note that unitary
transforms that do not mix orbitals from different bands, leave invariant the mean-field state that
is constructed by completely filling a single band. We use this freedom to re-Wannierize the mean-field orbitals within a given band.
The new Wannier orbitals are then given by
\begin{eqnarray}
\xi(j)&=&\frac{1}{N}\sum_{\alpha=-\frac{N-1}{2}}^{\frac{N-1}{2}}x_\alpha e^{i\phi^-_\alpha}e^{ik_\alpha j},\nonumber\\
\psi(j)&=&\frac{1}{N}\sum_{\alpha=-\frac{N-1}{2}}^{\frac{N-1}{2}}y_\alpha e^{i\phi^-_\alpha}e^{ik_\alpha j},\nonumber\\
\tilde\xi(j)&=&\frac{1}{N}\sum_{\alpha=-\frac{N-1}{2}}^{\frac{N-1}{2}}x_\alpha^* e^{i\phi^+_\alpha}e^{ik_\alpha j},\nonumber\\
\tilde\psi(j)&=&\frac{1}{N}\sum_{\alpha=-\frac{N-1}{2}}^{\frac{N-1}{2}}y_\alpha e^{i\phi^+_\alpha}e^{ik_\alpha j},\label{eq:defxipsi}
\end{eqnarray}
where the phases $\phi^\pm_\alpha$ have not been fixed yet, and may be adjusted to give optimal
results in whatever scheme is employed on top of mean field theory.  For our purposes, we used
$\phi^\pm_\alpha=0$, which gave a lower post-Hartree-Fock variational energy than other constant
choices. We did not systematically vary over momentum-dependent choices.

We use these orbitals to construct new
fermion creation operators $c_j^\dagger$. 
that are associated with Wannierized valence (even $j$) and conduction (odd $j$) band orbitals of
the mean-field Hamiltonian. In the position basis they read
\begin{eqnarray}
c_{2j}^\dagger&=&\sum_{l=1}^N\left[\xi(l-j)a_{2l-1}^\dagger+\psi(l-j)a_{2l}^\dagger\right],\nonumber\\
c_{2j-1}^\dagger&=&\sum_{l=1}^N\left[\tilde\psi(l-j)a_{2l-1}^\dagger-\tilde\xi(l-j)a_{2l}^\dagger\right].\label{eq:citoa}
\end{eqnarray}

It is useful to note that in terms of the parameters defined here, the expectation value of the
Hamiltonian with respect to the mean-field state is simply given by:
\begin{equation}
E_\text{mf}=-2NV\left(\frac{\tilde t^2-t^2}{V^2}+\frac{\Delta^2}{4}\right).\label{eq:mfenergy}
\end{equation}
In order to compare this result with perturbation theory, we wish to expand it in powers of $t/V$. As a first step, we expand $\Delta^2/V^2$ in $2\tilde t/\Delta= 1/z$. From
(\ref{eq:deltaandtt}) and (\ref{eq:diagas}) follows
\begin{align}
\frac{\Delta^2}{V^2}&=\frac{1}{N}\left[\sum_\alpha \left(1-\frac{1}{2z^2}\cos^2\frac{k_\alpha}{2}+\frac{3}{8z^4}\cos^4\frac{k_\alpha}{2}+\ldots\right)\right]^2\nonumber\\
&=\left(1-\frac{1}{4z^2}+\frac{9}{64z^4}+\ldots\right)^2\nonumber\\
&=1-\frac{1}{2z^2}+\frac{11}{32z^4}+\ldots
\end{align}
We can find $1/z$ in terms of $t/V$ with the help of (\ref{eq:findz}).
\begin{eqnarray}
\frac{2t}{V}&=&\frac{1}{Nz}\sum_\alpha\frac{\sin^2\frac{k_\alpha}{2}}{\sqrt{1+\frac{\cos^2 \frac{k_\alpha}{2} }{z^2}}}\nonumber\\
&=&\frac{1}{Nz}\sum_\alpha\left(\sin^2 \frac{k_\alpha}{2} -\frac{1}{2z^2}\sin^2 \frac{k_\alpha}{2}\cos^2 \frac{k_\alpha}{2}+\ldots\right) \nonumber\\
&=&\frac{1}{2z}\left[1-\frac{1}{8z^2}+\ldots\right] 
\end{eqnarray}
Solving the above for $1/z$, we find
\begin{equation}
\frac{1}{z}=\frac{4t}{V}+\frac{8t^3}{V^3}+\ldots
\end{equation}
Substituting these results into (\ref{eq:mfenergy}), we find
\begin{equation}
E_\text{mf}=-\frac{NV}{2}\left(1+\frac{4t^2}{V^2}-\frac{8t^4}{V^4}+\ldots\right).
\end{equation}
Comparing to the expansion of the exact result Eq.~(\ref{eq:E_0}), we see that the fourth order term 
of the Hartree-Fock solution is double what is should be. Correlations, included in the Coupled Cluster
state, are responsible for the proper value of the fourth order term.

\section{Matrix elements for two-body additive operators}
\label{app:matels}

Products of four $c$-operators, three of whose site indices have one parity (even/odd) and one whose site index is of opposite parity, necessarily change the number of sites of regular order by one. Such operators thus map $\left|m\right>$ onto a state in the orthogonal complement to the space spanned by the $\left|m'\right>$ $m'=0,1,\ldots,\lfloor N/2\rfloor$ basis. We therefore only have to consider operator products in which all four site indices have the same parity, or where two have one parity while the other two have the opposite parity.

\subsection{Operator products that increase $m$ by one}

Let us consider a two-body operator that increases $m$ by one. It necessarily consists of two
creation operators with odd site indices and two annihilation operators with even indices. They
fuse three disruptions (some of which could be length zero), that are separated by two single sites
of regular order into a single disruption, and grow the fused disruption by one site at each end.
Due to invariance of the state $\left|m\right>$ under translation by two lattice sites, we can
restrict our attention to products in which one of the creation operators has site index $1$. We
use three labels $l_1$, $l_2$ and $l_3$, corresponding to the number of atomic disruptions in each
of the three disruptions that are fused, to label these operator products. There are four separate
cases to consider, corresponding to how the fused disruption is positioned relative to the link
between sites $N$ and $1$. Note that because $N$ is odd, the initial Slater determinant on which the
operator product acts always hosts an odd number of particles in the space between the ends of the
eventual fused disruption.
\begin{enumerate}
\item $c_1^\dagger c_{4l_1+2}c_{4(l_1+l_2)+3}^\dagger c_{4(l_1+l_2+l_3+1)}$: Suppose the fused disruption starts at site $1$. The corresponding initial Slater determinant contains a site of regular order at site 1 (i.e. unoccupied),
followed by a disruption on sites $2$ to $4l_1+1$. A site of regular order at site $4l_1+2$ is occupied. Then follows a disruption from site $4l_1+3$ to site $4(l_1+l_2)+2$. A site of regular order at $4(l_1+l_2)+3$ is empty. Finally comes a disruption from site $4(l_1+l_2)+4$ to $4(l_1+l_2+l_3)+3$ and an occupied site of regular order at site $4(l_1+l_2+l_3)+3$.\\ 
\begin{align}
&\texttt{...| 0[0 1 0 1]1[1 0 1 0]0[0 1 0 1]1...}\to\nonumber\\
&\texttt{...|[1 0 1 0 1 0 1 0 1 0 1 0 1 0 1 0]...}\nonumber
\end{align}
Here and below \texttt{|} separates site $2N$ and site $1$ and.
Because of the initial presence of a particle on site $4l_1+2$ the 
operators $c_{4(l_1+l_2+l_3+1)}$ and 
$c_{4(l_1+l_2)+3}^\dagger$ each has to be anti-commuted past an odd number of fermion operators while $c_{4l_1+2}$ must be anti-commuted past an even number. This results in an overall sign of $1$.
\item $c_1^\dagger c_{2N-4(l_1+l_2+l_3)-2}c_{2N-4(l_2+l_3)-1}^\dagger c_{2N-4l_3}$: Suppose the fused disruption wraps around and ends at site $1$. Initially sites $1$ is empty while site $2N-4(l_1+l_2+l_3)-2$, just in front of the first unfused disruption is occupied. Site $2N-4(l_2+l_3)-1$ between the first and second unfused disruption is empty.\\ 
\begin{align}
&\texttt{|0 ...~1[1 0 1 0]0[0 1 0 1]1[1 0 1 0]|}\to\nonumber\\
&\texttt{|1]...[0 1 0 1 0 1 0 1 0 1 0 1 0 1 0 |}\nonumber
\end{align}
The odd number of particles hosted in sites $2$ to $2N-4(l_1+l_2+l_3)-2$ and the presence of a particle on site $2N-4(l_1+l_2+l_3)-2$ means that
$2c_{2N-4(l_2+l_3)-1}$ and $c_{2N-4l_3}$ must each be anti-commuted past an even number of particles, while $c_{2N-4(l_1+l_2+l_3)-2}$ is commuted past an odd number of particles. The overall sign that results is $-1$.
 \item $c_1^\dagger c_{4l_1+2}c_{2N-4(l_2+l_3)-1}^\dagger c_{2N-4l_3}$: One of the three fused disruptions fully wraps around past site $1$ (so that site $1$ is one of the fusion sites). In the initial Slater, site $1$, is therefore empty. Site $c_{4l_1+2}$, just after the insertion that wrapped past $1$, is initially filled. Site $2N-4(l_2+l_3)-1$, just in front of the remaining two insertions is initially empty.\\
\begin{align}
&\texttt{| 0[0 1 0 1]1~...~0[0 1 0 1]1[1 0 1 0]|}\to\nonumber\\
&\texttt{| 1 0 1 0 1 0]...[1 0 1 0 1 0 1 0 1 0 |}\nonumber
\end{align} 
Anti-commuting $c_{2N-4(l_2+l_3)-1}^\dagger$ and $c_{2N-4l_3}$ past odd number of particles in the space between the ends of the eventual fused insertion, and the particle on site $4l_1+2$ gives a sign of $1$. An even number of anti-commutations puts $c_{4l_1+2}$ in place. Thus the overall sign is $1$.
\item $c_1^\dagger c_{4l_1+2}c_{4(l_1+l_2)+3}^\dagger c_{2N-4l_3}$: Two of the three fused disruptions fully wrap around past site $1$. Site $1$, in front of the two insertions that wrap past $1$ is initially empty. Site $4l_1+2$ between the first and second wrapped insertion is initially filled. Site $4(l_1+l_2)+3$ just after the second wrapped insertion is initially empty. \\
\begin{align}
&\texttt{| 0[0 1 0 1]1[1 0 1 0]0~...~1[1 0 1 0]|}\to\nonumber\\
&\texttt{| 1 0 1 0 1 0 1 0 1 0 1]...[0 1 0 1 0 |}\nonumber\\
\end{align} 
Anti-commuting $c_{2N-4l_3}$ past the odd number of particles in the space between the ends of the eventual fused insertion, and past the particle on site $4l_1+2$, gives a sign of $1$. Anti-commuting $c_{4(l_1+l_2)+3}^\dagger$ past the particle on site $4l_1+2$, gives a sign of $-1$. Getting $ c_{4l_1+2}$ in place only requires anti-commuting it past the particles in one of the initial unfused disruptions, and thus gives a sign $1$. The overall sign is then $-1$.
\end{enumerate}
In each of the above cases, the number of Slater determinants in $\left|m\right>$ that fit the bill can be counted as follow. The three initially unfused insertions, together with the sites in between, in front of, or behind them, leaves $m-l_1-l_2-l_3$ atomic disruptions and $2N-4m-4$ sites of regular order to place in the space between the ends of the eventual fused insertion. There are ${2N-(3m+l_1+l_2+l_3)-4 \choose m-l_1-l_2-l_3}$ ways to do this. Each of these terms in $\left|m\right>$ is mapped onto a distinct term in $\left|m\right>$, with prefactor $\pm1$ as determined above. Thus
\begin{widetext}
\begin{align}
&\left<m+1\right|c_1^\dagger c_{4l_1+2}c_{4(l_1+l_2)+3}^\dagger c_{4(l_1+l_2+l_3+1)}\left|m\right>=-\left<m+1\right|c_1^\dagger c_{2N-4(l_1+l_2+l_3)-2}c_{2N-4(l_2+l_3)-1}^\dagger c_{2N-4l_3}\left|m\right>\nonumber\\
&=\left<m+1\right|c_1^\dagger c_{4l_1+2}c_{2N-4(l_2+l_3)-1}^\dagger c_{2N-4l_3}\left|m\right>=-\left<m+1\right|c_1^\dagger c_{4l_1+2}c_{4(l_1+l_2)+3}^\dagger c_{2N-4l_3}\left|m\right>=c^{1122}_{l_1,l_2,l_3},
\end{align}
\end{widetext}
where 
\begin{equation}
c^{1122}_{l_1,l_2,l_3}={2N-(3m+l_1+l_2+l_3)-4 \choose m-l_1-l_2-l_3}.
\end{equation}
The restriction that the initial unfused disruptions do not overlap, and that there should be no more than $m$ initial atomic disruptions, lead to the restrictions
\begin{align}
&l_1\in\left\{0,\min\left(\frac{N-1}{2}-1,m\right)\right\},\nonumber\\
&l_2\in\left\{0,\min\left(\frac{N-1}{2}-1-l_1,m-l_1\right)\right\},\nonumber\\
&l_3\in\left\{0,\min\left(\frac{N-1}{2}-1-l_1-l_2,m-l_1-l_2\right)\right\}.
\end{align}
Note that zero lengths are allowed for the unfused initial insertions. For instance, $l_1=l_2=l_3=0$ corresponds to a single atomic disruptions being created on the four sites 
from $1$ to $4$, if initially they are of regular order.\\
\centerline{\texttt{...|0101...} $\to$ \texttt{...|[1010]....}} 
Matrix elements of operators that reduce $m$ by one are obtained by taking the complex conjugates of the operators that we calculated here.

\subsection{A product in which all four site indices are distinct}
Such an operator product changes each of the Slater determinants $\left|s\right>=c^\dagger_{s_1}\ldots c^\dagger_{s_N}\left|0\right>$ that comprise $\left|m\right>$, into another Slater determinant
$\left|s'\right>=c^\dagger_{s_1'}\ldots c^\dagger_{s_N'}\left|0\right>$, up to a sign. Calculating 
the associated expectation value amounts to counting the number of these new Slater determinants $\left|s'\right>$ that are also part of $\left|m\right>$, 
or $\left|m\pm1\right>$, and keeping track of signs.

In each case, two disruptions are moved. There are four subclasses, determined by the directions in which the disruptions are moved (lefl-left, right-right, left-right, and right-left).
We use three integer labels $j$, $l_1$, and $l_2$ to enumerate the different operator products that yield a nonzero expectation value. 
\begin{enumerate}
\item Four odd site indices:
Without loss of generality, we can consider one creation operator to have site-index $1$. All other cases give the same answer, due to invariance of the states under consideration by two lattice sites. These operator products preserve the number of sites of regular order, and thus only connect $\left|m\right>$ to itself. The following mutually exclusive classes have non-zero $\left|m\right>$ expectation values.
\begin{enumerate}
\item $c_1^\dagger c_{4l_1+1} c_{2j+1}^\dagger c_{2j+1+4l_2}$: Suppose a Slater determinant contains a disruption starting on site $2$ and extending to site $4l_1+1$ 
and another present at site $2j+2$ and ending at site $2j+4l_2+1$. The operator product moves the first disruption one site to the left. This is achieved by 
anti-commuting $c_{4l_1+1}$ past all but the last fermion in the first disruption, thus accumulating a sign of $-1$. It also moves the second disruption one site to the left by anti-commuting $c_{2j+1+4l_2}$ past all but the right-most fermion in the second disruption, thus accumulating a sign $-1$.
\begin{align}
&\texttt{...| 0[0 1 0 1]...~0[0 1 0 1]...}\to\nonumber\\
&\texttt{...|[1 0 1 0]0~...[1 0 1 0]0~...}\nonumber
\end{align}
 The total sign is thus $1$.
\item $c_1^\dagger c_{2j+1-4l_1} c^\dagger_{2j+1} c_{2N-4l_2+1}$: Suppose a Slater determinant contains a disruption present on site $2j+1-4l_1$ and ending at site $2j$ 
and another present at site $2N+4l_2+1$ and ending at site $2N$. The operator product moves the first disruption one site to the right. This is achieved by 
anti-commuting $c_{2j+1}^\dagger$ past all the fermions in the first disruption, thus accumulating a sign of $1$. It also moves the second disruption one site to the right, so that its end wraps around to site $1$. It does so by anti-commuting $c_{2N-4l_2+1}$ past all the fermions that are not part of the right disruption. Since $N$ is odd, and disruptions contain an even number of fermions, this produces a sign $-1$.\\ 
\begin{align}
&\texttt{| 0~...[1 0 1 0]0~...[1 0 1 0]|} \to\nonumber\\
&\texttt{| 1]...~0[0 1 0 1]...~0[0 1 0 |}\nonumber
\end{align}
The total sign accumulated is then $-1$.
\item $c_1^\dagger c_{4l_1+1} c_{2j+1} c_{2j+1+4l_2}^\dagger$: Suppose a Slater determinant contains a disruption starting on site $2$ and extending to site $4l_1+1$ 
and another present at site $2j+1$ and ending at site $2j+4l_2$. The operator product moves the first disruption one site to the left. This is achieved by 
anti-commuting $c_{4l_1+1}$ past all but the right-most fermion in the first disruption, thus accumulating a sign of $-1$. It moves the second disruption one site to the right by anti-commuting $c^\dagger_{2j+1+4l_2}$ past all but the fermions in the second disruption, thus accumulating a sign $1$.\\ 
\begin{align}
&\texttt{...| 0[0 1 0 1]...[1 0 1 0]0~...}\to\nonumber\\
&\texttt{...|[1 0 1 0]0~...0[0 1 0 1]...}\nonumber
\end{align}
The total sign is thus $-1$.
\item $c_1^\dagger c^\dagger_{2j+1-4l_1} c_{2j+1} c_{2N-4l_2+1}$: Suppose a Slater determinant contains a disruption starting on site $2j+2-4l_1$ and extending to site $2j+1$ 
and another present at site $2N+4l_2+1$ and ending at site $2N$. The operator product moves the first disruption one site to the left. This is achieved by 
anti-commuting $c_{2j+1}$ past all but the right-most fermion in the first disruption, thus accumulating a sign of $-1$. It also moves the second disruption one site to the right, so that it end wraps around to site $1$. It does so by anti-commuting $c_{2N-4l_2+1}$ past all the fermions that are not part of the right disruption. This produces a sign $-1$. \\
\begin{align}
&\texttt{| 0~...~0[0 1 0 1]...[1 0 1 0]|} \to\nonumber\\
&\texttt{| 1]...[1 0 1 0]0~...~0[0 1 0 |}\nonumber
\end{align}
The total sign accumulated is then $1$.
\end{enumerate}
The number of Slater determinants in each sub-class is the same and can be computed as follows: the two disruptions that are moved together contain $l_1+l_2$ ``atomic'' (i.e. length= four sites) disruptions. Two sites of regular order (into which each disruption is moved) are also spoken for. This leaves $m-l_1-l_2$ atomic disruptions and $2N-2-4m$ sites
of regular order. The number of Slater determinants in a sub-class equals the number of ways we can arrange these atomic disruptions and sites of regular order in the space around the two moving disruptions. This space consists of two intervals, one of length $2j-4l_1-1$ and one of length $2N-2j-1-4l_2$. Let $q$ be the number of atomic insertions we assign to the latter interval. There are ${2N-2j-4l_2-1-3q \choose q}$ ways to do this. For each of these choices, there are ${2j-4l_1-1-3(m-q-l_2) \choose m-l_1-l_2-q}$ ways to arrange the remaining atomic disruptions within the interval of length $2j-4l_1-1$. The total number of ways is obtained by summing over $q$. 
\item A creation operator and an annihilation operator with odd site indices, and a creation operator and an annihilation operator with even site indices:
Without loss of generality, we can consider one creation operator to have site-index $1$. All other cases give the same answer, due to invariance of the states under consideration by two lattice sites. These operator products preserve the number of sites of regular order, and thus only connect $\left|m\right>$ to itself. The counting of Slater determinants proceeds in 
as in the case of four odd site indices, with only minor modifications.
\item Four even site indices:
Without loss of generality, we can consider one creation operator to have site-index $2$. All other cases give the same answer, due to invariance of the states under consideration by two lattice sites. These operator products preserve the number of sites of regular order, and thus only connect $\left|m\right>$ to itself. The counting of Slater determinants proceeds in 
as in the case of four odd site indices, with only minor modifications.
\end{enumerate}
The above considerations then lead to the following results. For operator products with four odd site indices:
\begin{align}
&\left<m\right|c_1^\dagger c_{4l_1+1} c_{2j+1}^\dagger c_{2j+1+4l_2}\left|m\right>\nonumber\\
=&-\left<m\right|c_1^\dagger c_{2j+1-4l_1} c^\dagger_{2j+1} c_{2N-4l_2+1}\left|m\right>\nonumber\\
=&-\left<m\right|c_1^\dagger c_{4l_1+1} c_{2j+1} c_{2j+1+4l_2}^\dagger\left|m\right>\nonumber\\
=&\left<m\right|c_1^\dagger c^\dagger_{2j+1-4l_1} c_{2j+1} c_{2N-4l_2+1}\left|m\right>\nonumber\\
=&c_{m,j,l_1,l_2}^{1111},
\end{align}
where
\begin{widetext}
\begin{equation}
c_{m,j,l_1,l_2}^{1111}=\sum_{q=\max\{0,\lceil m-l_2-\frac{j}{2}+\frac{1}{4}\rceil\}}^{\min\{m-l_1-l_2,\lfloor \frac{N-j}{2}-\frac{1}{4}-l_2\rfloor\}}{ 2N -2j -4l_2-1-3q \choose q}
{2j - l_1 -1 -3(m-q-l_2) \choose m-l_1 -l_2 -q }.
 \end{equation}
 \end{widetext}
The fact that the two disruptions are each at least four sites long, don't overlap, and jointly have length $4m$ sites, restricts the labels to the respective ranges 
\begin{align}
&j\in\{3,\ldots,N-3\},\nonumber\\
&l_1\in\left\{1,\min\left(\Big\lfloor\frac{j-1}{2}\Big\rfloor,m-1\right)\right\},\nonumber\\
&l_2\in\left\{1,\min\left(\Big\lfloor\frac{N-j-1}{2}\Big\rfloor,m-l_1\right)\right\}.
\end{align}

For an operator product with a creation operator and an annihilation operator with odd site indices, and a creation operator and an annihilation operator with even site indices
\begin{align}
&\left<m\right|c_1^\dagger c_{4l_1+1} c_{2j+2} c_{2j+2+4l_2}^\dagger\left|m\right>\nonumber\\
=&-\left<m\right|c_1^\dagger c_{2j+2-4l_1}^\dagger c_{2j+2} c_{2N-4l_2+1}\left|m\right>\nonumber\\
=&-\left<m\right|c_1^\dagger c_{4l_1+1} c_{2j+2}^\dagger c_{2j+2+4l_2}\left|m\right>\nonumber\\
=&\left<m\right|c_1^\dagger c_{2j+2-4l_1} c_{2j+2}^\dagger c_{2N-4l_2+1}\left|m\right>\nonumber\\
=&c_{m,j,l_1,l_2}^{1212},
\end{align}
where
\begin{widetext}
\begin{equation}
c_{m,j,l_1,l_2}^{1212}=\sum_{q=\max\{0,\lceil m-l_2-\frac{j}{2}\rceil\}}^{\min\{m-l_1-l_2,\lfloor \frac{N-j-1}{2}-l_2\rfloor\}}{ 2N -2j -4l_2-2-3q \choose q}
 {2j - l_1 -3(m-q-l_2) \choose m-l_1 -l_2 -q }.
\end{equation}
\end{widetext}
~\\
\break

Here
\begin{align}
&j\in\{2,\ldots,N-3\},\nonumber\\
&l_1\in\left\{1,\min\left(\Big\lfloor\frac{j}{2}\Big\rfloor,m-1\right)\right\},\nonumber\\
&l_2\in\left\{1,\min\left(\Big\lfloor\frac{N-j-1}{2}\Big\rfloor,m-l_1\right)\right\}.
\end{align}
For operator products with four even site indices:
\begin{align}
&-\left<m\right|c_2^\dagger c_{2j-4l_1+2} c_{2j+2}^\dagger c_{2N-4l_2+2}\left|m\right>\nonumber\\
=&\left<m\right|c_2^\dagger c_{4l_1+2} c^\dagger_{2j+2} c_{2j+4l_2+2}\left|m\right>\nonumber\\
=&\left<m\right|c_2^\dagger c_{2j-4l_1+2}^\dagger c_{2j+2} c_{2j-4l_2+2}\left|m\right>\nonumber\\
=&-\left<m\right|c_2^\dagger c_{4l_1+2} c_{2j+2} c_{2j+4l_2+2}^\dagger\left|m\right>\nonumber\\
=&c_{m,j,l_1,l_2}^{2222},
\end{align}
where $c_{m,j,l_1,l_2}^{2222}=c_{m,j,l_1,l_2}^{1111}$, i.e. the combinatorical factor in the last line of the above equation is the same as for the case of four odd site indices.)
The fact that the two disruptions are each at least four sites long, don't overlap, and jointly have length $4m$ sites, restricts the labels to the respective ranges 
\begin{align}
&j\in\{3,\ldots,N-3\},\nonumber\\
&l_1\in\left\{1,\min\left(\Big\lfloor\frac{j-1}{2}\Big\rfloor,m-1\right)\right\},\nonumber\\
&l_2\in\left\{1,\min\left(\Big\lfloor\frac{N-j-1}{2}\Big\rfloor,m-l_1\right)\right\}.
\end{align}

\subsection{A product in which one of the annihilation operators have the same site index as one of the creation operators}
In this case the two remaining operators must have site indices of the same parity, so that the only non-zero elements are between states with the same $m$.
Owing to invariance under translation by two lattice sites, we can focus on the cases where the repeated site index is either $1$ or $2$, and choose to consider
$\left<m\right|c_1c_1^\dagger \ldots\left|m\right>$ and $\left<m\right| c_2^\dagger c_2 \ldots\left|m\right>$. These operator products check that site $1$ or site $2$
are of regular order, and further moves one disruption to the left or right by one site.
The counting arguments are very similar to the case of four distinct site indices. We obtain
\begin{align}
&\left<m\right|c_1c_1^\dagger c_{2j+1}c_{2j+4l+1}^\dagger\left|m\right>\nonumber\\
=&-\left<m\right|c_1c_1^\dagger c_{2j+1}^\dagger c_{2j+4l+1}\left|m\right>\nonumber\\
=&c^{1111}_{m,j,0,l},
\end{align} 
with the restrictions on $j$ and $l$ being
\begin{align}
&j\in\{1,\ldots,N-3\},\nonumber\\
&l\in\left\{1,\min\left(\Big\lfloor\frac{N-j-1}{2}\Big\rfloor,m\right)\right\}.
\end{align}
Furthermore
\begin{align}
&\left<m\right|c_1c_1^\dagger c_{2j+2}^\dagger c_{2j+4l+2}\left|m\right>\nonumber\\
=&-\left<m\right|c_1c_1^\dagger c_{2j+2} c_{2j+4l+2}^\dagger\left|m\right>\nonumber\\
=&c^{1212}_{m,j,0,l},
\end{align} 
with 
\begin{align}
&j\in\{0,\ldots,N-3\},\nonumber\\
&l\in\left\{1,\min\left(\Big\lfloor\frac{N-j-1}{2}\Big\rfloor,m\right)\right\}.
\end{align}
Lastly
\begin{align}
&\left<m\right|c_2^\dagger c_2 c_{2j+2}^\dagger c_{2j+4l+2}\left|m\right>\nonumber\\
=&-\left<m\right|c_2^\dagger c_2 c_{2j+2} c_{2j+4l+2}^\dagger\left|m\right>\nonumber\\
=&c^{2222}_{m,j,0,l},
\end{align} 
with 
\begin{align}
&j\in\{1,\ldots,N-3\},\nonumber\\
&l\in\left\{1,\min\left(\Big\lfloor\frac{N-j-1}{2}\Big\rfloor,m\right)\right\}.
\end{align}

\subsection{A product with two pairs of creation and annihilation operators, where the members of each pair has the same site index}
Invariance under translation by two lattice sites allows us to focus on three classes:
\begin{equation}
\left<m\right|c_1c_1^\dagger c_{2j+1}c_{2j+1}^\dagger\left|m\right>=c^{1111}_{m,j,0,0}
\end{equation}
for $j\in\{1,\ldots,N-1\}$.
\begin{equation}
\left<m\right|c_1c_1^\dagger c_{2j+2}^\dagger c_{2j+2}\left|m\right>=c^{1212}_{m,j,0,0}
\end{equation}
for $j\in\{0,\ldots,N-1\}$.
\begin{equation}
\left<m\right|c_2^\dagger c_2 c_{2j+2}^\dagger c_{2j+2}\left|m\right>=c^{2222}_{m,j,0,0}
\end{equation}
for $j\in\{1,\ldots,N-1\}$.

\section{Hamiltonian matrix elements}
\label{app:ham}
All the ingredients necessary have now been collected and we can assemble them to express matrix elements of the Hamiltonian between $\left|m\right>$ states in terms of
the Wannier wave functions $\psi(l)$, $\xi(l)$, $\tilde\psi(l)$ and $\tilde\xi(l)$. 

The kinetic energy does not connect states with different $m$.
\begin{align}
&\left<m\right|-t\sum_{j=1}^{2N}\left(a_j^\dagger a_{j+1}+a_{j+1}^\dagger a_j \right)\left|m\right>\nonumber\\
=&Nt\Bigg[\frac{2N}{2N-3m}{2N-3m \choose m}X_0^{11}\nonumber\\
&-\sum_{l=0}^m\left(X_{2l}^{11}+X_{2l}^{22}\right){2N-3m-l-1 \choose m-l}\Bigg]+\text{c.c.}
\end{align}
where
\begin{eqnarray}
X^{11}_j&=&\sum_{l=1}^N \left[\tilde\psi(l)^*\tilde\xi(l-j)+\tilde\xi(l)^*\tilde\psi(l+1-j)\right],\nonumber\\
X^{22}_j&=&\sum_{l=1}^N \left[\xi(l)^*\psi(l-j)+\psi(l)^*\xi(l+1-j)\right].
\end{eqnarray}

The interaction term is more involved.
It connects states $\left|m\right>$ to itself and to $\left|m\pm1\right>$: 
\begin{equation}
\left<m\right|\sum_{j=1}^{2N}a_j^\dagger a_j a_{j+1}^\dagger a_{j+1}\left|m\right>=N\left(T^{1111}_m+T^{1212}_m+T^{2222}_m\right),
\end{equation} 
and
\begin{equation}
\left<m+1\right|\sum_{j=1}^{2N}a_j^\dagger a_j a_{j+1}^\dagger a_{j+1}\left|m\right>=NT^{1122}_m.
\end{equation} 
 Here 
 \begin{equation}
 T^{\sigma_1\sigma_2\sigma_3\sigma_4}_m= S^{\sigma_1\sigma_2\sigma_3\sigma_4}_{0,m}+S^{\sigma_1\sigma_2\sigma_3\sigma_4}_{1,m}+S^{\sigma_1\sigma_2\sigma_3\sigma_4}_{1,m}.
 \end{equation}
\begin{widetext}
 The $S_0$ contributions originate form four fermion operators that all have distinct Wannier orbital indices. For $\alpha=1111,\,1212,\,2222$,
 \begin{align}
 S_{0,m}^\alpha=\sum^\alpha c^\alpha_{m,j,l_1,l_2}\Big[&-W^\alpha(j+2l_2,2l_1,j)+W^\alpha(j+2l_2,j,2l_1)+W^\alpha(j,j+2l_2,2l_1)-W^\alpha(j,2l_1,j+2l_2)\nonumber\\
 &-W^\alpha(j,-2l_2,j-2l_1)+W^\alpha(j,j-2l_1,-2l_2)+W^\alpha(j-2l_1,j,-2l_2)-W^\alpha(j-2l_1,-2l_2,j)\Big]
 \end{align}
 For $\alpha=1122$,
 \begin{align}
 &S_{0,m}^{1122}=\sum^{1122} c^{1122}_{l_1,l_2,l_3}\Big[-W^{1122}(2(l_1+l_2)+1,2(l_1+l_2+l_3)+1,2l_1)+W^{1122}(2(l_1+l_2)+1,2l_1,2(l_1+l_2+l_3)+1)\nonumber\\
 &+W^{1122}(-2(l_2+l_3)-1,-2l_3-1,-2(l_1+l_2+l_3)-2)-W^{1122}(-2(l_2+l_3)-1,-2(l_1+l_2+l_3)-2,-2l_3-1)\nonumber\\
 &-W^{1122}(-2(l_2+l_3)-1,-2l_3-1,2l_1)+W^{1122}(-2(l_2+l_3)-1,2l_1,-2l_3-1)\nonumber\\
 &+W^{1122}(2(l_1+l_2)+1,-2l_3-1,2l_1)-W^{1122}(2(l_1+l_2)+1,2l_1,-2l_3-1)\Big].
 \end{align}
 The summations are
 \begin{align}
 &\sum^{1111}=\sum_{j=3}^{N-3}~\sum_{l_1=1}^{\min\{\lfloor\frac{j-1}{2}\rfloor,m-1\}}~\sum_{l_2=1}^{\min\{\lfloor\frac{N-j-1}{2}\rfloor,m-l_1\}}\nonumber\\
 &\sum^{1212}=\sum_{j=2}^{N-3}~\sum_{l_1=1}^{\min\{\lfloor\frac{j}{2}\rfloor,m-1\}}~\sum_{l_2=1}^{\min\{\lfloor\frac{N-j-1}{2}\rfloor,m-l_1\}}\nonumber\\
 &\sum^{2222}=\sum_{j=3}^{N-1}~\sum_{l_1=1}^{\min\{\lfloor\frac{j-1}{2}\rfloor,m-1\}}~\sum_{l_2=1}^{\min\{\lfloor\frac{N-j-1}{2}\rfloor,m-l_1\}}
 \end{align}
 and
 \begin{equation}
 \sum^{1122}=\sum_{l_1=0}^{\min\{\frac{N-1}{2}-1,m\}}~\sum_{l_2=0}^{\min\{\frac{N-1}{2}-1-l_1,m-l_1\}}~\sum_{l_3=0}^{\min\{\frac{N-1}{2}-1-l_1-l_2,m-l_1-l_2\}}.
 \end{equation}
 If the upper bound of a sum is lower than the lower bound, then the result is zero. 
 The $W^\alpha$'s read
\begin{align}
W^{1111}(j_1,j_2,j_3)=\sum_{l=1}^N \Big[&\tilde\psi(l)^*\tilde\xi(l-j_1)^*\tilde\xi(l-j_2)\tilde\psi(l-j_3)+ \tilde\xi(l)^*\tilde\psi(l+1-j_1)^*\tilde\psi(l+1-j_2)\tilde\xi(l-j_3)\Big],\nonumber\\
W^{1212}(j_1,j_2,j_3)=\sum_{l=1}^N \Bigg\{\Big[&\tilde\psi(l+1)^*\psi(l-j_1)^*+ \tilde\xi(l)^*\xi(l+1-j_1)^*\Big]\Big[\tilde\psi(l+1-j_2)\psi(l-j_3)+\tilde\xi(l-j_2)\xi(l+1-j_3)\Big]\nonumber\\
+\Big[&\tilde\psi(l)^*\psi(l-j_1)^*+\tilde\xi(l)^*\xi(l-j_1)^*\Big]\Big[\tilde\psi(l-j_2)\psi(l-j_3)+\tilde\xi(l-j_2)\xi(l-j_3)\Big]\Bigg\}.\nonumber\\
W^{2222}(j_1,j_2,j_3)=\sum_{l=1}^N \Big[&\psi(l)^*\xi(l+1-j_1)^*\xi(l+1-j_2)\psi(l-j_3)+ \xi(l)^*\psi(l-j_1)^*\psi(l-j_2)\xi(l-j_3)\Big].\nonumber\\
W^{1122}(j_1,j_2,j_3)=\sum_{l=1}^N \Big[&\tilde\psi(l)^*\tilde\xi(l-j_1)^*\psi(l-j_2)\xi(l-j_3)+ \tilde\xi(l)^*\tilde\psi(l+1-j_1)^*\xi(l+1-j_2)\psi(l-j_3)\Big],
\end{align}
The $S_1$ terms contain two Fermion $c$-operators with the same index. The $1122$ contribution is zero.
\begin{align}
&S_{1,m}^{1111}=\sum_{j=1}^{N-3}~\sum_{l=1}^{\min\{\lfloor\frac{N-1}{2}-\frac{j}{2}\rfloor,m\}}\left[V^{1111}(j,j+2l)+V^{1111}(j+2l,j)\right]\left[c^{1111}_{m,j,0,l}-{2N-3m-1-l \choose m-l}\right],\nonumber\\
&S_{1,m}^{2222}=\sum_{j=1}^{N-3}~\sum_{l=1}^{\min\{\lfloor\frac{N-1}{2}-\frac{j}{2}\rfloor,m\}}\left[V^{2222}(j,j+2l)+V^{2222}(j+2l,j)\right]c^{1111}_{m,j,0,l},
\end{align}
where 
\begin{equation}
V^\alpha(j_1,j_2)=W^\alpha(j_1,j_2,0)-W^\alpha(j_1,0,j_2)+W^\alpha(-j_1,-j_1,j_2-j_1)-W^\alpha(-j_1,j_2-j_1,-j_1).
\end{equation}
\begin{align}
S_{1,m}^{1212}=\sum_{j=0}^{N-3}~\sum_{l=1}^{\min\{\lfloor\frac{N-1}{2}-\frac{j}{2}\rfloor,m\}}\Bigg\{&\left[W^{1212}(j,0,j+2l)+W^{1212}(j+2l,0,j)\right]{2N-3m-1-l \choose m-l}\nonumber\\
&-\Big[W^{1212}(j,0,j+2l)+W^{1212}(j+2l,0,j)\nonumber\\
&+W^{1212}(-j-1,2l,-j-1)+W^{1212}(-j-2l-1,-2l,-j-2l-1)\Big]c_{m,j,0,l}^{1212}\Bigg\}.
\end{align}
The $S_2$ terms contain two pairs of Fermion $c$-operators with the same index. Again the $1122$ contribution is zero.
\begin{align}
S_{2,m}^{1111}=\sum_{j=1}^{N-1}&\left[W^{1111}(j,j,0)-W^{1111}(j,0,j)\right]\left[\frac{2N}{2N-3m}{2N-3m \choose m}-2{2N-1-3m \choose m}+c^{1111}_{m,j,0,0}\right]\nonumber\\
S_{2,m}^{1212}=\sum_{j=0}^{N-1}&W^{1212}(j,0,j) \left[{2N-1-3m \choose m}-c^{1212}_{m,j,0,0}\right] \nonumber\\
S_{2,m}^{2222}=\sum_{j=1}^{N-1}&\left[W^{2222}(j,j,0)-W^{2222}(j,0,j)\right]c^{1111}_{m,j,0,0}.
\end{align}
\end{widetext}

\section{A more general variational calculation}
\label{app:var}

In the main text, we showed that the exponential Coupled Cluster state $\left|\alpha\right>=\exp(\alpha T)
\left|\text{MF}\right>$ with $\alpha=(t/V)^2$ yields a variational energy that is accurate to
order $(t/V)^4$. In this appendix we investigate a more general state built from states
$\left|m\right>=\left(T\right)^m\left|\text{MF}\right>/m!$. We namely consider the ansatz
\begin{equation}
\left|\Psi\right>=\sum_{m=0}^{\lfloor N/2\rfloor} \Psi_m \left|m\right>.\label{eq:ansatz2}
\end{equation}
The simplest variational scheme is to treat the Wannierized single particle basis obtained in
mean-field theory as cast in stone, and to vary only the amplitudes $\Psi_m$ in (\ref{eq:ansatz}).
However, one may hope to obtain better results by also adjusting the single-particle basis as well. 
We have implemented the simplest version of this optimization by varying the parameter $z$, that defines 
the $c^\dagger_j$-orbitals through (\ref{eq:defx}-\ref{eq:defy}) and (\ref{eq:defxipsi}), along with the $\Psi_m$,
rather than fixing it by solving the mean-field equation (\ref{eq:findz}). 

\begin{center}
\begin{figure}
\includegraphics[width=0.99\columnwidth]{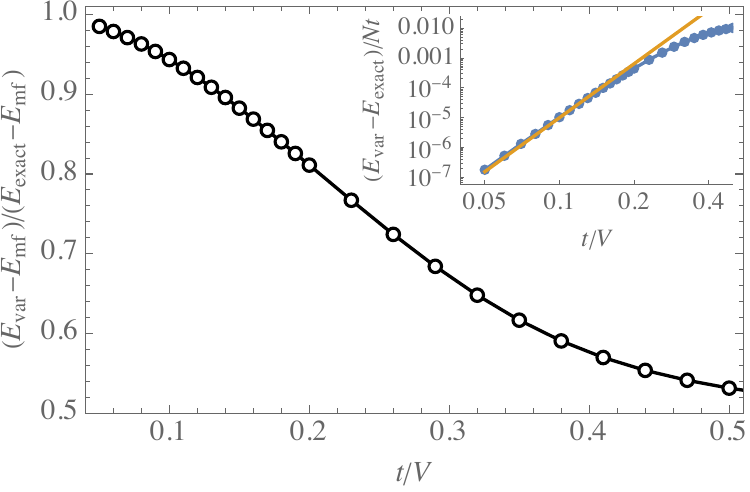}
\caption{\label{fig:energy2} {\bf Main panel:} The difference in ground state energy between 
mean-field theory (\ref{eq:mfenergy}) and the correlated ansatz (\ref{eq:ansatz2}) in units of 
the exact correlation energy. {\bf Inset:} The difference between the variational energy per particle
of the correlated ansatz (\ref{eq:ansatz2}) and the true ground state energy (\ref{eq:exact}) in
double log scale. The solid line corresponds to $10 (t/V)^6$. Results are shown for a system of
$N=41$ particles occupying $82$ sites.} 
\end{figure}
\end{center}

In Figure~\ref{fig:energy2}, we again plot the ratio between the correlation energy as estimated by
the more general ansatz (\ref{eq:ansatz2}), to the true correlation energy, as well as the
difference in energy per particle between the more general ansatz and the exact result. This should
be compared to Figure~\ref{fig:energy} in the main text. We see that the error in energy per
particle goes like $10(t/V)^6$ at small $t/V$, as compared to $19.5(t/V)^6$ for the simpler
exponential ansatz of the main text, giving a small improvement. 
At larger $t/V$, the more general ansatz more qualitatively capture the physics, as it can account for 
about 50\% of the correlation energy even close to the phase transition at $t/V=0.5$. 

\begin{center}
\begin{figure}
\includegraphics[width=0.99\columnwidth]{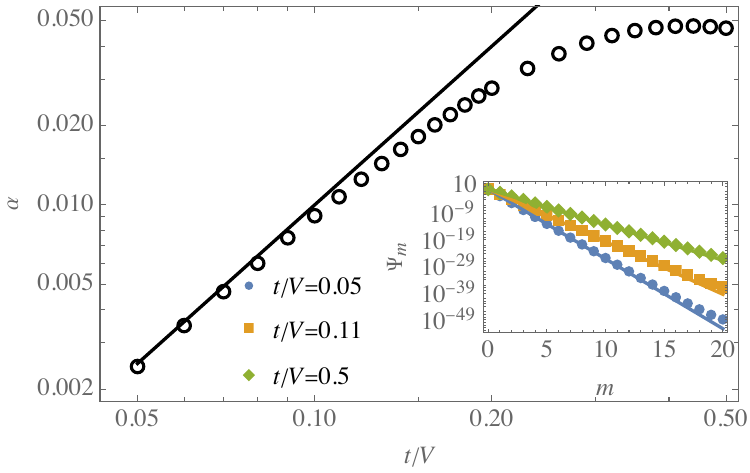}
\caption{\label{fig:alpha} {\bf Inset:} The optimal variational amplitudes $\Psi_m$
(\ref{eq:ansatz2}) versus number of disruptions $m$, plotted in logarithmic scale on the vertical
axis. Results for three values of $t/V$ are shown. Solid lines represent fits to the Coupled Cluster
formula $\Psi_m=N \alpha^m$. {\bf Main panel:} The extracted cluster strength $\alpha$ versus $t/V$
plotted in double log scale. The solid line corresponds to $\alpha=(t/V)^2$. Results are shown for a
system of $N=41$ particles occupying $82$ sites.} 
\end{figure}
\end{center}

In the inset of Figure~\ref{fig:alpha} we plot the state amplitudes $\Psi_m$ of the optimal
correlated states for various $t/V$, including ones outside the regime where the correlated ansatz
is quantitatively accurate. In all cases, we see that the amplitudes are very well approximated by a
function that is exponential in $m$. Thus the improved accuracy of more general ansatz
(\ref{eq:ansatz2}) does not come from abandoning the Coupled Clusters doubles form. Rather it is due
to a renormalization of $\alpha$ away from $(t/V)^2$ when this parameters becomes large, 
and to the modification of the mean-field parameter $z$. At small
$t/V$ where the correlated ansatz is intended to be quantitatively accurate, we find that varying
$z$ makes little difference. For instance, at $t/V=0.1$ the optimal $z$ is $2.40$ while the
mean-filed solution is $2.45$. At larger $t/V$ there is a more significant renormalization of $z$
(around $20\%$ close to the critical point), which leads to improved correlation energy estimates.
It should however be kept in mind that in this regime, the correlated ansatz provides at best a
qualitative description anyway, because the periodicity of the covariance matrix spectrum is
not correctly obtained. In the main panel of Figure~\ref{fig:alpha}, we plot the $\alpha$
values versus $t/V$ that we extracted from our improved variational calculation. In the regime $t\lesssim0.1V$,
where the ansatz is quantitatively accurate, we still find that $\alpha\simeq (t/V)^2$, but at larger
$t/V$ it is suppressed below this quadratic growth. 

\section{Explicit expression for $\zeta_0(\alpha)$}
\label{app:zeta}
In Sec.~\ref{sec:parent}, we expressed the covariance matrix associated with the state $\left|\alpha\right>$ in terms of the 
real root $\zeta_0$ of $1-\zeta-\alpha^2\zeta^4$ that is closest to the origin. 
Using Ferrari's method to solve this depressed quartic equation we obtain:
\begin{equation}
\zeta_0(\alpha)=\frac{1}{2\alpha^{2/3}}\left[\sqrt{\frac{2}{u(\alpha)}-u(\alpha)^2}-u(\alpha)\right],
\end{equation}
where $u(\alpha)$ takes the value
\begin{align}
&\sqrt{\left(\frac{\sqrt{1+\frac{256}{27}\alpha^2}+1}{2}\right)^{1/3}-\left(\frac{\sqrt{1+\frac{256}{27}\alpha^2}-1}{2}\right)^{1/3}}.
\end{align}


\begin{thebibliography}{50}%
\makeatletter
\providecommand \@ifxundefined [1]{%
 \@ifx{#1\undefined}
}%
\providecommand \@ifnum [1]{%
 \ifnum #1\expandafter \@firstoftwo
 \else \expandafter \@secondoftwo
 \fi
}%
\providecommand \@ifx [1]{%
 \ifx #1\expandafter \@firstoftwo
 \else \expandafter \@secondoftwo
 \fi
}%
\providecommand \natexlab [1]{#1}%
\providecommand \enquote  [1]{``#1''}%
\providecommand \bibnamefont  [1]{#1}%
\providecommand \bibfnamefont [1]{#1}%
\providecommand \citenamefont [1]{#1}%
\providecommand \href@noop [0]{\@secondoftwo}%
\providecommand \href [0]{\begingroup \@sanitize@url \@href}%
\providecommand \@href[1]{\@@startlink{#1}\@@href}%
\providecommand \@@href[1]{\endgroup#1\@@endlink}%
\providecommand \@sanitize@url [0]{\catcode `\\12\catcode `\$12\catcode
  `\&12\catcode `\#12\catcode `\^12\catcode `\_12\catcode `\%12\relax}%
\providecommand \@@startlink[1]{}%
\providecommand \@@endlink[0]{}%
\providecommand \url  [0]{\begingroup\@sanitize@url \@url }%
\providecommand \@url [1]{\endgroup\@href {#1}{\urlprefix }}%
\providecommand \urlprefix  [0]{URL }%
\providecommand \Eprint [0]{\href }%
\providecommand \doibase [0]{https://doi.org/}%
\providecommand \selectlanguage [0]{\@gobble}%
\providecommand \bibinfo  [0]{\@secondoftwo}%
\providecommand \bibfield  [0]{\@secondoftwo}%
\providecommand \translation [1]{[#1]}%
\providecommand \BibitemOpen [0]{}%
\providecommand \bibitemStop [0]{}%
\providecommand \bibitemNoStop [0]{.\EOS\space}%
\providecommand \EOS [0]{\spacefactor3000\relax}%
\providecommand \BibitemShut  [1]{\csname bibitem#1\endcsname}%
\let\auto@bib@innerbib\@empty
\bibitem [{\citenamefont {Giuliani}\ and\ \citenamefont
  {Vignale}(2005)}]{Giuliani}%
  \BibitemOpen
  \bibfield  {author} {\bibinfo {author} {\bibfnamefont {G.~F.}\ \bibnamefont
  {Giuliani}}\ and\ \bibinfo {author} {\bibfnamefont {G.}~\bibnamefont
  {Vignale}},\ }\href@noop {} {\emph {\bibinfo {title} {Quantum theory of the
  electron liquid}}}\ (\bibinfo  {publisher} {Cambridge},\ \bibinfo {year}
  {2005})\BibitemShut {NoStop}%
\bibitem [{\citenamefont {L{\"o}wdin}(1955)}]{Lowdin1955}%
  \BibitemOpen
  \bibfield  {author} {\bibinfo {author} {\bibfnamefont {P.-O.}\ \bibnamefont
  {L{\"o}wdin}},\ }\bibfield  {title} {\bibinfo {title} {Quantum theory of
  many-particle systems. {I.} {Physical} interpretations by means of density
  matrices, natural spin-orbitals, and convergence problems in the method of
  configurational interaction},\ }\href
  {https://doi.org/10.1103/PhysRev.97.1474} {\bibfield  {journal} {\bibinfo
  {journal} {Physical Review}\ }\textbf {\bibinfo {volume} {97}},\ \bibinfo
  {pages} {1474} (\bibinfo {year} {1955})}\BibitemShut {NoStop}%
\bibitem [{\citenamefont {Davidson}(1972)}]{Davidson1972}%
  \BibitemOpen
  \bibfield  {author} {\bibinfo {author} {\bibfnamefont {E.~R.}\ \bibnamefont
  {Davidson}},\ }\bibfield  {title} {\bibinfo {title} {Properties and uses of
  natural orbitals},\ }\href {https://doi.org/10.1103/RevModPhys.44.451}
  {\bibfield  {journal} {\bibinfo  {journal} {Rev. Mod. Phys.}\ }\textbf
  {\bibinfo {volume} {44}},\ \bibinfo {pages} {451} (\bibinfo {year}
  {1972})}\BibitemShut {NoStop}%
\bibitem [{\citenamefont {Bruus}\ and\ \citenamefont
  {Flensberg}(2015)}]{Flensberg}%
  \BibitemOpen
  \bibfield  {author} {\bibinfo {author} {\bibfnamefont {H.}~\bibnamefont
  {Bruus}}\ and\ \bibinfo {author} {\bibfnamefont {K.}~\bibnamefont
  {Flensberg}},\ }\href@noop {} {\emph {\bibinfo {title} {Many-body quantum
  theory in condensed matter physics}}}\ (\bibinfo  {publisher} {Oxford},\
  \bibinfo {year} {2015})\BibitemShut {NoStop}%
\bibitem [{\citenamefont {Giamarchi}(2003)}]{Giamarchi2003}%
  \BibitemOpen
  \bibfield  {author} {\bibinfo {author} {\bibfnamefont {T.}~\bibnamefont
  {Giamarchi}},\ }\href@noop {} {\emph {\bibinfo {title} {Quantum physics in
  one dimension}}}\ (\bibinfo  {publisher} {Oxford},\ \bibinfo {year}
  {2003})\BibitemShut {NoStop}%
\bibitem [{\citenamefont {Olsen}(2011)}]{Olsen2011}%
  \BibitemOpen
  \bibfield  {author} {\bibinfo {author} {\bibfnamefont {J.}~\bibnamefont
  {Olsen}},\ }\bibfield  {title} {\bibinfo {title} {The {CASSCF} method: {A}
  perspective and commentary},\ }\href
  {https://doi.org/https://doi.org/10.1002/qua.23107} {\bibfield  {journal}
  {\bibinfo  {journal} {International Journal of Quantum Chemistry}\ }\textbf
  {\bibinfo {volume} {111}},\ \bibinfo {pages} {3267} (\bibinfo {year}
  {2011})}\BibitemShut {NoStop}%
\bibitem [{\citenamefont {He}\ and\ \citenamefont {Lu}(2014)}]{He2014}%
  \BibitemOpen
  \bibfield  {author} {\bibinfo {author} {\bibfnamefont {R.-Q.}\ \bibnamefont
  {He}}\ and\ \bibinfo {author} {\bibfnamefont {Z.-Y.}\ \bibnamefont {Lu}},\
  }\bibfield  {title} {\bibinfo {title} {Quantum renormalization groups based
  on natural orbitals},\ }\href {https://doi.org/10.1103/PhysRevB.89.085108}
  {\bibfield  {journal} {\bibinfo  {journal} {Phys. Rev. B}\ }\textbf {\bibinfo
  {volume} {89}},\ \bibinfo {pages} {085108} (\bibinfo {year}
  {2014})}\BibitemShut {NoStop}%
\bibitem [{\citenamefont {Debertolis}\ \emph {et~al.}(2021)\citenamefont
  {Debertolis}, \citenamefont {Florens},\ and\ \citenamefont
  {Snyman}}]{Debertolis2021}%
  \BibitemOpen
  \bibfield  {author} {\bibinfo {author} {\bibfnamefont {M.}~\bibnamefont
  {Debertolis}}, \bibinfo {author} {\bibfnamefont {S.}~\bibnamefont
  {Florens}},\ and\ \bibinfo {author} {\bibfnamefont {I.}~\bibnamefont
  {Snyman}},\ }\bibfield  {title} {\bibinfo {title} {Few-body nature of {Kondo}
  correlated ground states},\ }\href
  {https://doi.org/10.1103/PhysRevB.103.235166} {\bibfield  {journal} {\bibinfo
   {journal} {Physical Review B}\ }\textbf {\bibinfo {volume} {103}},\ \bibinfo
  {pages} {235166} (\bibinfo {year} {2021})}\BibitemShut {NoStop}%
\bibitem [{\citenamefont {Snyman}(2023)}]{Snyman2023}%
  \BibitemOpen
  \bibfield  {author} {\bibinfo {author} {\bibfnamefont {I.}~\bibnamefont
  {Snyman}},\ }\bibfield  {title} {\bibinfo {title} {Structure of
  quasiparticles in a local fermi liquid},\ }\href
  {https://doi.org/10.1103/PhysRevB.108.205120} {\bibfield  {journal} {\bibinfo
   {journal} {Phys. Rev. B}\ }\textbf {\bibinfo {volume} {108}},\ \bibinfo
  {pages} {205120} (\bibinfo {year} {2023})}\BibitemShut {NoStop}%
\bibitem [{\citenamefont {Bravyi}\ and\ \citenamefont
  {Gosset}(2017)}]{Bravyi2017}%
  \BibitemOpen
  \bibfield  {author} {\bibinfo {author} {\bibfnamefont {S.}~\bibnamefont
  {Bravyi}}\ and\ \bibinfo {author} {\bibfnamefont {D.}~\bibnamefont
  {Gosset}},\ }\bibfield  {title} {\bibinfo {title} {Complexity of quantum
  impurity problems},\ }\href {https://doi.org/10.1007/s00220-017-2976-9}
  {\bibfield  {journal} {\bibinfo  {journal} {Communications in Mathematical
  Physics}\ }\textbf {\bibinfo {volume} {356}},\ \bibinfo {pages} {451}
  (\bibinfo {year} {2017})}\BibitemShut {NoStop}%
\bibitem [{\citenamefont {Debertolis}\ \emph {et~al.}(2022)\citenamefont
  {Debertolis}, \citenamefont {Snyman},\ and\ \citenamefont
  {Florens}}]{Debertolis2022}%
  \BibitemOpen
  \bibfield  {author} {\bibinfo {author} {\bibfnamefont {M.}~\bibnamefont
  {Debertolis}}, \bibinfo {author} {\bibfnamefont {I.}~\bibnamefont {Snyman}},\
  and\ \bibinfo {author} {\bibfnamefont {S.}~\bibnamefont {Florens}},\
  }\bibfield  {title} {\bibinfo {title} {Simulating realistic screening clouds
  around quantum impurities: {Role} of spatial anisotropy and disorder},\
  }\href {https://doi.org/10.1103/PhysRevB.106.125115} {\bibfield  {journal}
  {\bibinfo  {journal} {Phys. Rev. B}\ }\textbf {\bibinfo {volume} {106}},\
  \bibinfo {pages} {125115} (\bibinfo {year} {2022})}\BibitemShut {NoStop}%
\bibitem [{\citenamefont {N{\'u}{\~n}ez-Fern{\'a}ndez}\ \emph
  {et~al.}(2025)\citenamefont {N{\'u}{\~n}ez-Fern{\'a}ndez}, \citenamefont
  {Debertolis},\ and\ \citenamefont {Florens}}]{Nunez2025}%
  \BibitemOpen
  \bibfield  {author} {\bibinfo {author} {\bibfnamefont {Y.}~\bibnamefont
  {N{\'u}{\~n}ez-Fern{\'a}ndez}}, \bibinfo {author} {\bibfnamefont
  {M.}~\bibnamefont {Debertolis}},\ and\ \bibinfo {author} {\bibfnamefont
  {S.}~\bibnamefont {Florens}},\ }\href {https://arxiv.org/abs/2503.13706}
  {\bibinfo {title} {Resolving space-time structures of quantum impurities with
  a numerically-exact algorithm using few-body revealing}} (\bibinfo {year}
  {2025}),\ \Eprint {https://arxiv.org/abs/2503.13706} {arXiv:2503.13706
  [cond-mat.str-el]} \BibitemShut {NoStop}%
\bibitem [{\citenamefont {Bera}\ \emph {et~al.}(2015)\citenamefont {Bera},
  \citenamefont {Schomerus}, \citenamefont {Heidrich-Meisner},\ and\
  \citenamefont {Bardarson}}]{Bera2015}%
  \BibitemOpen
  \bibfield  {author} {\bibinfo {author} {\bibfnamefont {S.}~\bibnamefont
  {Bera}}, \bibinfo {author} {\bibfnamefont {H.}~\bibnamefont {Schomerus}},
  \bibinfo {author} {\bibfnamefont {F.}~\bibnamefont {Heidrich-Meisner}},\ and\
  \bibinfo {author} {\bibfnamefont {J.~H.}\ \bibnamefont {Bardarson}},\
  }\bibfield  {title} {\bibinfo {title} {Many-body localization characterized
  from a one-particle perspective},\ }\href
  {https://doi.org/10.1103/PhysRevLett.115.046603} {\bibfield  {journal}
  {\bibinfo  {journal} {Phys. Rev. Lett.}\ }\textbf {\bibinfo {volume} {115}},\
  \bibinfo {pages} {046603} (\bibinfo {year} {2015})}\BibitemShut {NoStop}%
\bibitem [{\citenamefont {Thamm}\ \emph {et~al.}(2022)\citenamefont {Thamm},
  \citenamefont {Radhakrishnan}, \citenamefont {Barghathi}, \citenamefont
  {Rosenow},\ and\ \citenamefont {Del~Maestro}}]{Thamm2022}%
  \BibitemOpen
  \bibfield  {author} {\bibinfo {author} {\bibfnamefont {M.}~\bibnamefont
  {Thamm}}, \bibinfo {author} {\bibfnamefont {H.}~\bibnamefont
  {Radhakrishnan}}, \bibinfo {author} {\bibfnamefont {H.}~\bibnamefont
  {Barghathi}}, \bibinfo {author} {\bibfnamefont {B.}~\bibnamefont {Rosenow}},\
  and\ \bibinfo {author} {\bibfnamefont {A.}~\bibnamefont {Del~Maestro}},\
  }\bibfield  {title} {\bibinfo {title} {One-particle entanglement for
  one-dimensional spinless fermions after an interaction quantum quench},\
  }\href {https://doi.org/10.1103/PhysRevB.106.165116} {\bibfield  {journal}
  {\bibinfo  {journal} {Phys. Rev. B}\ }\textbf {\bibinfo {volume} {106}},\
  \bibinfo {pages} {165116} (\bibinfo {year} {2022})}\BibitemShut {NoStop}%
\bibitem [{\citenamefont {Vanhala}\ and\ \citenamefont
  {Ojanen}(2024)}]{Vanhala2024}%
  \BibitemOpen
  \bibfield  {author} {\bibinfo {author} {\bibfnamefont {T.~I.}\ \bibnamefont
  {Vanhala}}\ and\ \bibinfo {author} {\bibfnamefont {T.}~\bibnamefont
  {Ojanen}},\ }\bibfield  {title} {\bibinfo {title} {Complexity of fermionic
  states},\ }\href {https://doi.org/10.1103/PhysRevResearch.6.023178}
  {\bibfield  {journal} {\bibinfo  {journal} {Phys. Rev. Res.}\ }\textbf
  {\bibinfo {volume} {6}},\ \bibinfo {pages} {023178} (\bibinfo {year}
  {2024})}\BibitemShut {NoStop}%
\bibitem [{\citenamefont {Des~Cloizeaux}(1966)}]{Cloizeaux1966}%
  \BibitemOpen
  \bibfield  {author} {\bibinfo {author} {\bibfnamefont {J.}~\bibnamefont
  {Des~Cloizeaux}},\ }\bibfield  {title} {\bibinfo {title} {A soluble
  {F}ermi‐gas model. {V}alidity of transformations of the {B}ogoliubov
  type},\ }\href {https://doi.org/10.1063/1.1704899} {\bibfield  {journal}
  {\bibinfo  {journal} {Journal of Mathematical Physics}\ }\textbf {\bibinfo
  {volume} {7}},\ \bibinfo {pages} {2136} (\bibinfo {year} {1966})}\BibitemShut
  {NoStop}%
\bibitem [{\citenamefont {Yang}\ and\ \citenamefont {Yang}(1966)}]{Yang1966}%
  \BibitemOpen
  \bibfield  {author} {\bibinfo {author} {\bibfnamefont {C.~N.}\ \bibnamefont
  {Yang}}\ and\ \bibinfo {author} {\bibfnamefont {C.~P.}\ \bibnamefont
  {Yang}},\ }\bibfield  {title} {\bibinfo {title} {One-dimensional chain of
  anisotropic spin-spin interactions. {II}. {P}roperties of the ground-state
  energy per lattice site for an infinite system},\ }\href
  {https://doi.org/10.1103/PhysRev.150.327} {\bibfield  {journal} {\bibinfo
  {journal} {Phys. Rev.}\ }\textbf {\bibinfo {volume} {150}},\ \bibinfo {pages}
  {327} (\bibinfo {year} {1966})}\BibitemShut {NoStop}%
\bibitem [{\citenamefont {White}(1992)}]{White1992}%
  \BibitemOpen
  \bibfield  {author} {\bibinfo {author} {\bibfnamefont {S.~R.}\ \bibnamefont
  {White}},\ }\bibfield  {title} {\bibinfo {title} {Density matrix formulation
  for quantum renormalization groups},\ }\href
  {https://doi.org/10.1103/PhysRevLett.69.2863} {\bibfield  {journal} {\bibinfo
   {journal} {Phys. Rev. Lett.}\ }\textbf {\bibinfo {volume} {69}},\ \bibinfo
  {pages} {2863} (\bibinfo {year} {1992})}\BibitemShut {NoStop}%
\bibitem [{\citenamefont {Karrasch}\ and\ \citenamefont
  {Moore}(2012)}]{Karrasch2012}%
  \BibitemOpen
  \bibfield  {author} {\bibinfo {author} {\bibfnamefont {C.}~\bibnamefont
  {Karrasch}}\ and\ \bibinfo {author} {\bibfnamefont {J.~E.}\ \bibnamefont
  {Moore}},\ }\bibfield  {title} {\bibinfo {title} {Luttinger liquid physics
  from the infinite-system density matrix renormalization group},\ }\href
  {https://doi.org/10.1103/PhysRevB.86.155156} {\bibfield  {journal} {\bibinfo
  {journal} {Phys. Rev. B}\ }\textbf {\bibinfo {volume} {86}},\ \bibinfo
  {pages} {155156} (\bibinfo {year} {2012})}\BibitemShut {NoStop}%
\bibitem [{\citenamefont {Jimbo}\ \emph {et~al.}(1992)\citenamefont {Jimbo},
  \citenamefont {Miki}, \citenamefont {Miwa},\ and\ \citenamefont
  {Nakayashiki}}]{Jimbo1992}%
  \BibitemOpen
  \bibfield  {author} {\bibinfo {author} {\bibfnamefont {M.}~\bibnamefont
  {Jimbo}}, \bibinfo {author} {\bibfnamefont {K.}~\bibnamefont {Miki}},
  \bibinfo {author} {\bibfnamefont {T.}~\bibnamefont {Miwa}},\ and\ \bibinfo
  {author} {\bibfnamefont {A.}~\bibnamefont {Nakayashiki}},\ }\bibfield
  {title} {\bibinfo {title} {Correlation functions of the {$XXZ$} model for
  {$\Delta<-1$}},\ }\href
  {https://doi.org/https://doi.org/10.1016/0375-9601(92)91128-E} {\bibfield
  {journal} {\bibinfo  {journal} {Physics Letters A}\ }\textbf {\bibinfo
  {volume} {168}},\ \bibinfo {pages} {256} (\bibinfo {year}
  {1992})}\BibitemShut {NoStop}%
\bibitem [{\citenamefont {G{\"o}hmann}\ \emph {et~al.}(2005)\citenamefont
  {G{\"o}hmann}, \citenamefont {Kl{\"u}mper},\ and\ \citenamefont
  {Seel}}]{Gohmann2005}%
  \BibitemOpen
  \bibfield  {author} {\bibinfo {author} {\bibfnamefont {F.}~\bibnamefont
  {G{\"o}hmann}}, \bibinfo {author} {\bibfnamefont {A.}~\bibnamefont
  {Kl{\"u}mper}},\ and\ \bibinfo {author} {\bibfnamefont {A.}~\bibnamefont
  {Seel}},\ }\bibfield  {title} {\bibinfo {title} {Integral representation of
  the density matrix of the {$XXZ$} chain at finite temperatures},\ }\href
  {https://doi.org/10.1088/0305-4470/38/9/001} {\bibfield  {journal} {\bibinfo
  {journal} {Journal of Physics A: Mathematical and General}\ }\textbf
  {\bibinfo {volume} {38}},\ \bibinfo {pages} {1833} (\bibinfo {year}
  {2005})}\BibitemShut {NoStop}%
\bibitem [{\citenamefont {Kitanine}\ \emph {et~al.}(2000)\citenamefont
  {Kitanine}, \citenamefont {Maillet},\ and\ \citenamefont
  {Terras}}]{Kitanine2000}%
  \BibitemOpen
  \bibfield  {author} {\bibinfo {author} {\bibfnamefont {N.}~\bibnamefont
  {Kitanine}}, \bibinfo {author} {\bibfnamefont {J.}~\bibnamefont {Maillet}},\
  and\ \bibinfo {author} {\bibfnamefont {V.}~\bibnamefont {Terras}},\
  }\bibfield  {title} {\bibinfo {title} {Correlation functions of the {$XXZ$}
  heisenberg spin-{$1/2$} chain in a magnetic field},\ }\href
  {https://doi.org/https://doi.org/10.1016/S0550-3213(99)00619-7} {\bibfield
  {journal} {\bibinfo  {journal} {Nuclear Physics B}\ }\textbf {\bibinfo
  {volume} {567}},\ \bibinfo {pages} {554} (\bibinfo {year}
  {2000})}\BibitemShut {NoStop}%
\bibitem [{\citenamefont {Babenko}\ \emph {et~al.}(2021)\citenamefont
  {Babenko}, \citenamefont {G\"ohmann}, \citenamefont {Kozlowski},
  \citenamefont {Sirker},\ and\ \citenamefont {Suzuki}}]{Babenko2021}%
  \BibitemOpen
  \bibfield  {author} {\bibinfo {author} {\bibfnamefont {C.}~\bibnamefont
  {Babenko}}, \bibinfo {author} {\bibfnamefont {F.}~\bibnamefont {G\"ohmann}},
  \bibinfo {author} {\bibfnamefont {K.~K.}\ \bibnamefont {Kozlowski}}, \bibinfo
  {author} {\bibfnamefont {J.}~\bibnamefont {Sirker}},\ and\ \bibinfo {author}
  {\bibfnamefont {J.}~\bibnamefont {Suzuki}},\ }\bibfield  {title} {\bibinfo
  {title} {Exact real-time longitudinal correlation functions of the massive
  {$XXZ$} chain},\ }\href {https://doi.org/10.1103/PhysRevLett.126.210602}
  {\bibfield  {journal} {\bibinfo  {journal} {Phys. Rev. Lett.}\ }\textbf
  {\bibinfo {volume} {126}},\ \bibinfo {pages} {210602} (\bibinfo {year}
  {2021})}\BibitemShut {NoStop}%
\bibitem [{\citenamefont {Gebhard}\ \emph {et~al.}(2022)\citenamefont
  {Gebhard}, \citenamefont {Bauerbach},\ and\ \citenamefont
  {Legeza}}]{Gebhard2022}%
  \BibitemOpen
  \bibfield  {author} {\bibinfo {author} {\bibfnamefont {F.}~\bibnamefont
  {Gebhard}}, \bibinfo {author} {\bibfnamefont {K.}~\bibnamefont {Bauerbach}},\
  and\ \bibinfo {author} {\bibfnamefont {O.}~\bibnamefont {Legeza}},\
  }\bibfield  {title} {\bibinfo {title} {Accurate localization of
  {K}osterlitz-{T}houless-type quantum phase transitions for one-dimensional
  spinless fermions},\ }\href {https://doi.org/10.1103/PhysRevB.106.205133}
  {\bibfield  {journal} {\bibinfo  {journal} {Phys. Rev. B}\ }\textbf {\bibinfo
  {volume} {106}},\ \bibinfo {pages} {205133} (\bibinfo {year}
  {2022})}\BibitemShut {NoStop}%
\bibitem [{\citenamefont {Bartlett}\ and\ \citenamefont
  {Musia\l{}}(2007)}]{Bartlett2007}%
  \BibitemOpen
  \bibfield  {author} {\bibinfo {author} {\bibfnamefont {R.~J.}\ \bibnamefont
  {Bartlett}}\ and\ \bibinfo {author} {\bibfnamefont {M.}~\bibnamefont
  {Musia\l{}}},\ }\bibfield  {title} {\bibinfo {title} {Coupled-cluster theory
  in quantum chemistry},\ }\href {https://doi.org/10.1103/RevModPhys.79.291}
  {\bibfield  {journal} {\bibinfo  {journal} {Rev. Mod. Phys.}\ }\textbf
  {\bibinfo {volume} {79}},\ \bibinfo {pages} {291} (\bibinfo {year}
  {2007})}\BibitemShut {NoStop}%
\bibitem [{\citenamefont {Hirata}\ \emph {et~al.}(2004)\citenamefont {Hirata},
  \citenamefont {Podeszwa}, \citenamefont {Tobita},\ and\ \citenamefont
  {Bartlett}}]{Hirata2004}%
  \BibitemOpen
  \bibfield  {author} {\bibinfo {author} {\bibfnamefont {S.}~\bibnamefont
  {Hirata}}, \bibinfo {author} {\bibfnamefont {R.}~\bibnamefont {Podeszwa}},
  \bibinfo {author} {\bibfnamefont {M.}~\bibnamefont {Tobita}},\ and\ \bibinfo
  {author} {\bibfnamefont {R.~J.}\ \bibnamefont {Bartlett}},\ }\bibfield
  {title} {\bibinfo {title} {Coupled-cluster singles and doubles for extended
  systems},\ }\href {https://doi.org/10.1063/1.1637577} {\bibfield  {journal}
  {\bibinfo  {journal} {The Journal of Chemical Physics}\ }\textbf {\bibinfo
  {volume} {120}},\ \bibinfo {pages} {2581} (\bibinfo {year} {2004})},\ \Eprint
  {https://arxiv.org/abs/https://pubs.aip.org/aip/jcp/article-pdf/120/6/2581/19316695/2581\_1\_online.pdf}
  {https://pubs.aip.org/aip/jcp/article-pdf/120/6/2581/19316695/2581\_1\_online.pdf}
  \BibitemShut {NoStop}%
\bibitem [{\citenamefont {Bishop}\ and\ \citenamefont
  {L\"uhrmann}(1978)}]{Bishop1978}%
  \BibitemOpen
  \bibfield  {author} {\bibinfo {author} {\bibfnamefont {R.~F.}\ \bibnamefont
  {Bishop}}\ and\ \bibinfo {author} {\bibfnamefont {K.~H.}\ \bibnamefont
  {L\"uhrmann}},\ }\bibfield  {title} {\bibinfo {title} {Electron correlations:
  {I}. ground-state results in the high-density regime},\ }\href
  {https://doi.org/10.1103/PhysRevB.17.3757} {\bibfield  {journal} {\bibinfo
  {journal} {Phys. Rev. B}\ }\textbf {\bibinfo {volume} {17}},\ \bibinfo
  {pages} {3757} (\bibinfo {year} {1978})}\BibitemShut {NoStop}%
\bibitem [{\citenamefont {Shepherd}\ \emph {et~al.}(2012)\citenamefont
  {Shepherd}, \citenamefont {Gr\"uneis}, \citenamefont {Booth}, \citenamefont
  {Kresse},\ and\ \citenamefont {Alavi}}]{Shepherd2012}%
  \BibitemOpen
  \bibfield  {author} {\bibinfo {author} {\bibfnamefont {J.~J.}\ \bibnamefont
  {Shepherd}}, \bibinfo {author} {\bibfnamefont {A.}~\bibnamefont {Gr\"uneis}},
  \bibinfo {author} {\bibfnamefont {G.~H.}\ \bibnamefont {Booth}}, \bibinfo
  {author} {\bibfnamefont {G.}~\bibnamefont {Kresse}},\ and\ \bibinfo {author}
  {\bibfnamefont {A.}~\bibnamefont {Alavi}},\ }\bibfield  {title} {\bibinfo
  {title} {Convergence of many-body wave-function expansions using a plane-wave
  basis: From homogeneous electron gas to solid state systems},\ }\href
  {https://doi.org/10.1103/PhysRevB.86.035111} {\bibfield  {journal} {\bibinfo
  {journal} {Phys. Rev. B}\ }\textbf {\bibinfo {volume} {86}},\ \bibinfo
  {pages} {035111} (\bibinfo {year} {2012})}\BibitemShut {NoStop}%
\bibitem [{\citenamefont {Georgiou}\ \emph {et~al.}(2024)\citenamefont
  {Georgiou}, \citenamefont {Rousochatzakis}, \citenamefont {Farnell},
  \citenamefont {Richter},\ and\ \citenamefont {Bishop}}]{Georgio2024}%
  \BibitemOpen
  \bibfield  {author} {\bibinfo {author} {\bibfnamefont {M.}~\bibnamefont
  {Georgiou}}, \bibinfo {author} {\bibfnamefont {I.}~\bibnamefont
  {Rousochatzakis}}, \bibinfo {author} {\bibfnamefont {D.~J.~J.}\ \bibnamefont
  {Farnell}}, \bibinfo {author} {\bibfnamefont {J.}~\bibnamefont {Richter}},\
  and\ \bibinfo {author} {\bibfnamefont {R.~F.}\ \bibnamefont {Bishop}},\
  }\bibfield  {title} {\bibinfo {title} {Spin-{$S$} {K}itaev-{H}eisenberg model
  on the honeycomb lattice: A high-order treatment via the many-body coupled
  cluster method},\ }\href {https://doi.org/10.1103/PhysRevResearch.6.033168}
  {\bibfield  {journal} {\bibinfo  {journal} {Phys. Rev. Res.}\ }\textbf
  {\bibinfo {volume} {6}},\ \bibinfo {pages} {033168} (\bibinfo {year}
  {2024})}\BibitemShut {NoStop}%
\bibitem [{\citenamefont {Cirac}\ \emph {et~al.}(2021)\citenamefont {Cirac},
  \citenamefont {P\'erez-Garc\'{\i}a}, \citenamefont {Schuch},\ and\
  \citenamefont {Verstraete}}]{Cirac2021}%
  \BibitemOpen
  \bibfield  {author} {\bibinfo {author} {\bibfnamefont {J.~I.}\ \bibnamefont
  {Cirac}}, \bibinfo {author} {\bibfnamefont {D.}~\bibnamefont
  {P\'erez-Garc\'{\i}a}}, \bibinfo {author} {\bibfnamefont {N.}~\bibnamefont
  {Schuch}},\ and\ \bibinfo {author} {\bibfnamefont {F.}~\bibnamefont
  {Verstraete}},\ }\bibfield  {title} {\bibinfo {title} {Matrix product states
  and projected entangled pair states: Concepts, symmetries, theorems},\ }\href
  {https://doi.org/10.1103/RevModPhys.93.045003} {\bibfield  {journal}
  {\bibinfo  {journal} {Rev. Mod. Phys.}\ }\textbf {\bibinfo {volume} {93}},\
  \bibinfo {pages} {045003} (\bibinfo {year} {2021})}\BibitemShut {NoStop}%
\bibitem [{\citenamefont {Sharma}\ and\ \citenamefont
  {Alavi}(2015)}]{Sharma2015}%
  \BibitemOpen
  \bibfield  {author} {\bibinfo {author} {\bibfnamefont {S.}~\bibnamefont
  {Sharma}}\ and\ \bibinfo {author} {\bibfnamefont {A.}~\bibnamefont {Alavi}},\
  }\bibfield  {title} {\bibinfo {title} {Multireference linearized coupled
  cluster theory for strongly correlated systems using matrix product states},\
  }\href {https://doi.org/10.1063/1.4928643} {\bibfield  {journal} {\bibinfo
  {journal} {The Journal of Chemical Physics}\ }\textbf {\bibinfo {volume}
  {143}},\ \bibinfo {pages} {102815} (\bibinfo {year} {2015})}\BibitemShut
  {NoStop}%
\bibitem [{\citenamefont {Veis}\ \emph {et~al.}(2016)\citenamefont {Veis},
  \citenamefont {Antal{\'\i}k}, \citenamefont {Brabec}, \citenamefont {Neese},
  \citenamefont {Legeza},\ and\ \citenamefont {Pittner}}]{Veis2016}%
  \BibitemOpen
  \bibfield  {author} {\bibinfo {author} {\bibfnamefont {L.}~\bibnamefont
  {Veis}}, \bibinfo {author} {\bibfnamefont {A.}~\bibnamefont {Antal{\'\i}k}},
  \bibinfo {author} {\bibfnamefont {J.}~\bibnamefont {Brabec}}, \bibinfo
  {author} {\bibfnamefont {F.}~\bibnamefont {Neese}}, \bibinfo {author}
  {\bibfnamefont {{\"O}.}~\bibnamefont {Legeza}},\ and\ \bibinfo {author}
  {\bibfnamefont {J.}~\bibnamefont {Pittner}},\ }\bibfield  {title} {\bibinfo
  {title} {Coupled cluster method with single and double excitations tailored
  by matrix product state wave functions},\ }\href
  {https://doi.org/10.1021/acs.jpclett.6b01908} {\bibfield  {journal} {\bibinfo
   {journal} {The Journal of Physical Chemistry Letters}\ }\textbf {\bibinfo
  {volume} {7}},\ \bibinfo {pages} {4072} (\bibinfo {year} {2016})}\BibitemShut
  {NoStop}%
\bibitem [{\citenamefont {Ramirez-Pastor}\ \emph {et~al.}(1999)\citenamefont
  {Ramirez-Pastor}, \citenamefont {Eggarter}, \citenamefont {Pereyra},\ and\
  \citenamefont {Riccardo}}]{Ramirez-Pastor1999}%
  \BibitemOpen
  \bibfield  {author} {\bibinfo {author} {\bibfnamefont {A.~J.}\ \bibnamefont
  {Ramirez-Pastor}}, \bibinfo {author} {\bibfnamefont {T.~P.}\ \bibnamefont
  {Eggarter}}, \bibinfo {author} {\bibfnamefont {V.~D.}\ \bibnamefont
  {Pereyra}},\ and\ \bibinfo {author} {\bibfnamefont {J.~L.}\ \bibnamefont
  {Riccardo}},\ }\bibfield  {title} {\bibinfo {title} {Statistical
  thermodynamics and transport of linear adsorbates},\ }\href
  {https://doi.org/10.1103/PhysRevB.59.11027} {\bibfield  {journal} {\bibinfo
  {journal} {Phys. Rev. B}\ }\textbf {\bibinfo {volume} {59}},\ \bibinfo
  {pages} {11027} (\bibinfo {year} {1999})}\BibitemShut {NoStop}%
\bibitem [{\citenamefont {Castelnovo}\ \emph {et~al.}(2005)\citenamefont
  {Castelnovo}, \citenamefont {Chamon}, \citenamefont {Mudry},\ and\
  \citenamefont {Pujol}}]{Castelnovo2005}%
  \BibitemOpen
  \bibfield  {author} {\bibinfo {author} {\bibfnamefont {C.}~\bibnamefont
  {Castelnovo}}, \bibinfo {author} {\bibfnamefont {C.}~\bibnamefont {Chamon}},
  \bibinfo {author} {\bibfnamefont {C.}~\bibnamefont {Mudry}},\ and\ \bibinfo
  {author} {\bibfnamefont {P.}~\bibnamefont {Pujol}},\ }\bibfield  {title}
  {\bibinfo {title} {From quantum mechanics to classical statistical physics:
  Generalized {R}okhsar--{K}ivelson {H}amiltonians and the ``{S}tochastic
  {M}atrix {F}orm'' decomposition},\ }\href
  {https://doi.org/https://doi.org/10.1016/j.aop.2005.01.006} {\bibfield
  {journal} {\bibinfo  {journal} {Annals of Physics}\ }\textbf {\bibinfo
  {volume} {318}},\ \bibinfo {pages} {316} (\bibinfo {year}
  {2005})}\BibitemShut {NoStop}%
\bibitem [{\citenamefont {Rokhsar}\ and\ \citenamefont
  {Kivelson}(1988)}]{Rokhsar1988}%
  \BibitemOpen
  \bibfield  {author} {\bibinfo {author} {\bibfnamefont {D.~S.}\ \bibnamefont
  {Rokhsar}}\ and\ \bibinfo {author} {\bibfnamefont {S.~A.}\ \bibnamefont
  {Kivelson}},\ }\bibfield  {title} {\bibinfo {title} {Superconductivity and
  the quantum hard-core dimer gas},\ }\href
  {https://doi.org/10.1103/PhysRevLett.61.2376} {\bibfield  {journal} {\bibinfo
   {journal} {Phys. Rev. Lett.}\ }\textbf {\bibinfo {volume} {61}},\ \bibinfo
  {pages} {2376} (\bibinfo {year} {1988})}\BibitemShut {NoStop}%
\bibitem [{\citenamefont {Fannes}\ \emph {et~al.}(1992)\citenamefont {Fannes},
  \citenamefont {Nachtergaele},\ and\ \citenamefont {Werner}}]{Fannes1992}%
  \BibitemOpen
  \bibfield  {author} {\bibinfo {author} {\bibfnamefont {M.}~\bibnamefont
  {Fannes}}, \bibinfo {author} {\bibfnamefont {B.}~\bibnamefont
  {Nachtergaele}},\ and\ \bibinfo {author} {\bibfnamefont {R.~F.}\ \bibnamefont
  {Werner}},\ }\bibfield  {title} {\bibinfo {title} {Finitely correlated states
  on quantum spin chains},\ }\href {https://doi.org/10.1007/BF02099178}
  {\bibfield  {journal} {\bibinfo  {journal} {Communications in Mathematical
  Physics}\ }\textbf {\bibinfo {volume} {144}},\ \bibinfo {pages} {443}
  (\bibinfo {year} {1992})}\BibitemShut {NoStop}%
\bibitem [{\citenamefont {Fern{\'a}ndez-Gonz{\'a}lez}\ \emph
  {et~al.}(2015)\citenamefont {Fern{\'a}ndez-Gonz{\'a}lez}, \citenamefont
  {Schuch}, \citenamefont {Wolf}, \citenamefont {Cirac},\ and\ \citenamefont
  {P{\'e}rez-Garc{\'\i}a}}]{Fernandez-Gonzalez2015}%
  \BibitemOpen
  \bibfield  {author} {\bibinfo {author} {\bibfnamefont {C.}~\bibnamefont
  {Fern{\'a}ndez-Gonz{\'a}lez}}, \bibinfo {author} {\bibfnamefont
  {N.}~\bibnamefont {Schuch}}, \bibinfo {author} {\bibfnamefont {M.~M.}\
  \bibnamefont {Wolf}}, \bibinfo {author} {\bibfnamefont {J.~I.}\ \bibnamefont
  {Cirac}},\ and\ \bibinfo {author} {\bibfnamefont {D.}~\bibnamefont
  {P{\'e}rez-Garc{\'\i}a}},\ }\bibfield  {title} {\bibinfo {title} {Frustration
  free gapless {H}amiltonians for matrix product states},\ }\href
  {https://doi.org/10.1007/s00220-014-2173-z} {\bibfield  {journal} {\bibinfo
  {journal} {Communications in Mathematical Physics}\ }\textbf {\bibinfo
  {volume} {333}},\ \bibinfo {pages} {299} (\bibinfo {year}
  {2015})}\BibitemShut {NoStop}%
\bibitem [{\citenamefont {Fishman}\ \emph {et~al.}(2022)\citenamefont
  {Fishman}, \citenamefont {White},\ and\ \citenamefont
  {Stoudenmire}}]{ITensor2022}%
  \BibitemOpen
  \bibfield  {author} {\bibinfo {author} {\bibfnamefont {M.}~\bibnamefont
  {Fishman}}, \bibinfo {author} {\bibfnamefont {S.~R.}\ \bibnamefont {White}},\
  and\ \bibinfo {author} {\bibfnamefont {E.~M.}\ \bibnamefont {Stoudenmire}},\
  }\bibfield  {title} {\bibinfo {title} {The {ITensor} software library for
  tensor network calculations},\ }\href
  {https://doi.org/10.21468/SciPostPhysCodeb.4} {\bibfield  {journal} {\bibinfo
   {journal} {SciPost Phys. Codebases}\ ,\ \bibinfo {pages} {4}} (\bibinfo
  {year} {2022})}\BibitemShut {NoStop}%
\bibitem [{\citenamefont {Barghathi}\ \emph {et~al.}(2019)\citenamefont
  {Barghathi}, \citenamefont {Casiano-Diaz},\ and\ \citenamefont
  {Del~Maestro}}]{Barghathi2019}%
  \BibitemOpen
  \bibfield  {author} {\bibinfo {author} {\bibfnamefont {H.}~\bibnamefont
  {Barghathi}}, \bibinfo {author} {\bibfnamefont {E.}~\bibnamefont
  {Casiano-Diaz}},\ and\ \bibinfo {author} {\bibfnamefont {A.}~\bibnamefont
  {Del~Maestro}},\ }\bibfield  {title} {\bibinfo {title} {Operationally
  accessible entanglement of one-dimensional spinless fermions},\ }\href
  {https://doi.org/10.1103/PhysRevA.100.022324} {\bibfield  {journal} {\bibinfo
   {journal} {Phys. Rev. A}\ }\textbf {\bibinfo {volume} {100}},\ \bibinfo
  {pages} {022324} (\bibinfo {year} {2019})}\BibitemShut {NoStop}%
\bibitem [{\citenamefont {Kohno}\ \emph {et~al.}(2010)\citenamefont {Kohno},
  \citenamefont {Arikawa}, \citenamefont {Sato},\ and\ \citenamefont
  {Sakai}}]{Kohno2010}%
  \BibitemOpen
  \bibfield  {author} {\bibinfo {author} {\bibfnamefont {M.}~\bibnamefont
  {Kohno}}, \bibinfo {author} {\bibfnamefont {M.}~\bibnamefont {Arikawa}},
  \bibinfo {author} {\bibfnamefont {J.}~\bibnamefont {Sato}},\ and\ \bibinfo
  {author} {\bibfnamefont {K.}~\bibnamefont {Sakai}},\ }\bibfield  {title}
  {\bibinfo {title} {Spectral properties of interacting one-dimensional
  spinless fermions},\ }\href {https://doi.org/10.1143/JPSJ.79.043707}
  {\bibfield  {journal} {\bibinfo  {journal} {Journal of the Physical Society
  of Japan}\ }\textbf {\bibinfo {volume} {79}},\ \bibinfo {pages} {043707}
  (\bibinfo {year} {2010})},\ \Eprint
  {https://arxiv.org/abs/https://journals.jps.jp/doi/pdf/10.1143/JPSJ.79.043707}
  {https://journals.jps.jp/doi/pdf/10.1143/JPSJ.79.043707} \BibitemShut
  {NoStop}%
\bibitem [{\citenamefont {Fratini}\ \emph {et~al.}(2004)\citenamefont
  {Fratini}, \citenamefont {Valenzuela},\ and\ \citenamefont
  {Baeriswyl}}]{Fratini2004}%
  \BibitemOpen
  \bibfield  {author} {\bibinfo {author} {\bibfnamefont {S.}~\bibnamefont
  {Fratini}}, \bibinfo {author} {\bibfnamefont {B.}~\bibnamefont
  {Valenzuela}},\ and\ \bibinfo {author} {\bibfnamefont {D.}~\bibnamefont
  {Baeriswyl}},\ }\bibfield  {title} {\bibinfo {title} {Incipient quantum
  melting of the one-dimensional {W}igner lattice},\ }\href
  {https://doi.org/https://doi.org/10.1016/j.synthmet.2003.10.030} {\bibfield
  {journal} {\bibinfo  {journal} {Synthetic Metals}\ }\textbf {\bibinfo
  {volume} {141}},\ \bibinfo {pages} {193} (\bibinfo {year} {2004})},\ \bibinfo
  {note} {michael J. Rice Memorial Festschrift}\BibitemShut {NoStop}%
\bibitem [{\citenamefont {Mayr}\ and\ \citenamefont {Horsch}(2006)}]{Mayr2006}%
  \BibitemOpen
  \bibfield  {author} {\bibinfo {author} {\bibfnamefont {M.}~\bibnamefont
  {Mayr}}\ and\ \bibinfo {author} {\bibfnamefont {P.}~\bibnamefont {Horsch}},\
  }\bibfield  {title} {\bibinfo {title} {Domain-wall excitations and optical
  conductivity in one-dimensional {W}igner lattices},\ }\href
  {https://doi.org/10.1103/PhysRevB.73.195103} {\bibfield  {journal} {\bibinfo
  {journal} {Phys. Rev. B}\ }\textbf {\bibinfo {volume} {73}},\ \bibinfo
  {pages} {195103} (\bibinfo {year} {2006})}\BibitemShut {NoStop}%
\bibitem [{\citenamefont {Mahyaeh}\ \emph {et~al.}(2022)\citenamefont
  {Mahyaeh}, \citenamefont {K\"ohler}, \citenamefont {Black-Schaffer},\ and\
  \citenamefont {Kantian}}]{Mahyaeh2022}%
  \BibitemOpen
  \bibfield  {author} {\bibinfo {author} {\bibfnamefont {I.}~\bibnamefont
  {Mahyaeh}}, \bibinfo {author} {\bibfnamefont {T.}~\bibnamefont {K\"ohler}},
  \bibinfo {author} {\bibfnamefont {A.~M.}\ \bibnamefont {Black-Schaffer}},\
  and\ \bibinfo {author} {\bibfnamefont {A.}~\bibnamefont {Kantian}},\
  }\bibfield  {title} {\bibinfo {title} {Superconducting pairing from repulsive
  interactions of fermions in a flat-band system},\ }\href
  {https://doi.org/10.1103/PhysRevB.106.125155} {\bibfield  {journal} {\bibinfo
   {journal} {Phys. Rev. B}\ }\textbf {\bibinfo {volume} {106}},\ \bibinfo
  {pages} {125155} (\bibinfo {year} {2022})}\BibitemShut {NoStop}%
\bibitem [{\citenamefont {Yang}(1989)}]{Yang1989}%
  \BibitemOpen
  \bibfield  {author} {\bibinfo {author} {\bibfnamefont {C.~N.}\ \bibnamefont
  {Yang}},\ }\bibfield  {title} {\bibinfo {title} {\ensuremath{\eta} pairing
  and off-diagonal long-range order in a {H}ubbard model},\ }\href
  {https://doi.org/10.1103/PhysRevLett.63.2144} {\bibfield  {journal} {\bibinfo
   {journal} {Phys. Rev. Lett.}\ }\textbf {\bibinfo {volume} {63}},\ \bibinfo
  {pages} {2144} (\bibinfo {year} {1989})}\BibitemShut {NoStop}%
\bibitem [{\citenamefont {Gotta}\ \emph {et~al.}(2022)\citenamefont {Gotta},
  \citenamefont {Mazza}, \citenamefont {Simon},\ and\ \citenamefont
  {Roux}}]{Lorenzo2022}%
  \BibitemOpen
  \bibfield  {author} {\bibinfo {author} {\bibfnamefont {L.}~\bibnamefont
  {Gotta}}, \bibinfo {author} {\bibfnamefont {L.}~\bibnamefont {Mazza}},
  \bibinfo {author} {\bibfnamefont {P.}~\bibnamefont {Simon}},\ and\ \bibinfo
  {author} {\bibfnamefont {G.}~\bibnamefont {Roux}},\ }\bibfield  {title}
  {\bibinfo {title} {Exact many-body scars based on pairs or multimers in a
  chain of spinless fermions},\ }\href
  {https://doi.org/10.1103/PhysRevB.106.235147} {\bibfield  {journal} {\bibinfo
   {journal} {Phys. Rev. B}\ }\textbf {\bibinfo {volume} {106}},\ \bibinfo
  {pages} {235147} (\bibinfo {year} {2022})}\BibitemShut {NoStop}%
\bibitem [{\citenamefont {Moessner}\ and\ \citenamefont
  {Sondhi}(2001)}]{Moessner2001}%
  \BibitemOpen
  \bibfield  {author} {\bibinfo {author} {\bibfnamefont {R.}~\bibnamefont
  {Moessner}}\ and\ \bibinfo {author} {\bibfnamefont {S.~L.}\ \bibnamefont
  {Sondhi}},\ }\bibfield  {title} {\bibinfo {title} {Resonating valence bond
  phase in the triangular lattice quantum dimer model},\ }\href
  {https://doi.org/10.1103/PhysRevLett.86.1881} {\bibfield  {journal} {\bibinfo
   {journal} {Phys. Rev. Lett.}\ }\textbf {\bibinfo {volume} {86}},\ \bibinfo
  {pages} {1881} (\bibinfo {year} {2001})}\BibitemShut {NoStop}%
\bibitem [{\citenamefont {Moudgalya}\ \emph {et~al.}(2018)\citenamefont
  {Moudgalya}, \citenamefont {Rachel}, \citenamefont {Bernevig},\ and\
  \citenamefont {Regnault}}]{Moudgalya2018}%
  \BibitemOpen
  \bibfield  {author} {\bibinfo {author} {\bibfnamefont {S.}~\bibnamefont
  {Moudgalya}}, \bibinfo {author} {\bibfnamefont {S.}~\bibnamefont {Rachel}},
  \bibinfo {author} {\bibfnamefont {B.~A.}\ \bibnamefont {Bernevig}},\ and\
  \bibinfo {author} {\bibfnamefont {N.}~\bibnamefont {Regnault}},\ }\bibfield
  {title} {\bibinfo {title} {Exact excited states of nonintegrable models},\
  }\href {https://doi.org/10.1103/PhysRevB.98.235155} {\bibfield  {journal}
  {\bibinfo  {journal} {Phys. Rev. B}\ }\textbf {\bibinfo {volume} {98}},\
  \bibinfo {pages} {235155} (\bibinfo {year} {2018})}\BibitemShut {NoStop}%
\bibitem [{\citenamefont {Meyer}(2000)}]{Meyer2000}%
  \BibitemOpen
  \bibfield  {author} {\bibinfo {author} {\bibfnamefont {C.}~\bibnamefont
  {Meyer}},\ }\href@noop {} {\emph {\bibinfo {title} {Matrix analysis and
  applied linear algebra}}}\ (\bibinfo  {publisher} {Society for Applied and
  Industrial Mathematics},\ \bibinfo {address} {Philedelphia PA},\ \bibinfo
  {year} {2000})\BibitemShut {NoStop}%
\bibitem [{\citenamefont {Menczer}\ \emph {et~al.}(2024)\citenamefont
  {Menczer}, \citenamefont {Kap\'as}, \citenamefont {Werner},\ and\
  \citenamefont {Legeza}}]{Menczer2024}%
  \BibitemOpen
  \bibfield  {author} {\bibinfo {author} {\bibfnamefont {A.}~\bibnamefont
  {Menczer}}, \bibinfo {author} {\bibfnamefont {K.}~\bibnamefont {Kap\'as}},
  \bibinfo {author} {\bibfnamefont {M.~A.}\ \bibnamefont {Werner}},\ and\
  \bibinfo {author} {\bibfnamefont {O.}~\bibnamefont {Legeza}},\ }\bibfield
  {title} {\bibinfo {title} {Two-dimensional quantum lattice models via mode
  optimized hybrid {CPU-GPU} density matrix renormalization group method},\
  }\href {https://doi.org/10.1103/PhysRevB.109.195148} {\bibfield  {journal}
  {\bibinfo  {journal} {Phys. Rev. B}\ }\textbf {\bibinfo {volume} {109}},\
  \bibinfo {pages} {195148} (\bibinfo {year} {2024})}\BibitemShut {NoStop}%
\bibitem [{\citenamefont {Hannukainen}\ \emph {et~al.}(2024)\citenamefont
  {Hannukainen}, \citenamefont {Mart\'{\i}nez}, \citenamefont {Bardarson},\
  and\ \citenamefont {Kvorning}}]{Hannukainen2024}%
  \BibitemOpen
  \bibfield  {author} {\bibinfo {author} {\bibfnamefont {J.~D.}\ \bibnamefont
  {Hannukainen}}, \bibinfo {author} {\bibfnamefont {M.~F.}\ \bibnamefont
  {Mart\'{\i}nez}}, \bibinfo {author} {\bibfnamefont {J.~H.}\ \bibnamefont
  {Bardarson}},\ and\ \bibinfo {author} {\bibfnamefont {T.~K.}\ \bibnamefont
  {Kvorning}},\ }\bibfield  {title} {\bibinfo {title} {Interacting local
  topological markers: A one-particle density matrix approach for
  characterizing the topology of interacting and disordered states},\ }\href
  {https://doi.org/10.1103/PhysRevResearch.6.L032045} {\bibfield  {journal}
  {\bibinfo  {journal} {Phys. Rev. Res.}\ }\textbf {\bibinfo {volume} {6}},\
  \bibinfo {pages} {L032045} (\bibinfo {year} {2024})}\BibitemShut {NoStop}%
\end{thebibliography}
\end{document}